\documentclass[sigconf, screen]{acmart}

\AtBeginDocument{%
  }

\setcopyright{none} 
\copyrightyear{2026}
\acmYear{2026}

\acmConference[]{2026}{}{}

\usepackage{makecell}
\usepackage{graphicx}
\usepackage{textcomp}
\usepackage{xcolor}
\usepackage{fancyhdr}
\usepackage{hyperref}
\usepackage{caption}
\usepackage{subcaption}
\usepackage{tikz}
\usepackage{enumitem}

\usepackage[ruled,linesnumbered]{algorithm2e}
\usepackage{amsmath}

\usepackage{listings}
\usepackage{xcolor}
\definecolor{codegreen}{rgb}{0.2, 0.6, 0.2}
\definecolor{codepurple}{rgb}{0.58, 0.0, 0.83}
\definecolor{codeorange}{rgb}{0.8, 0.4, 0.0}
\definecolor{codeblue}{rgb}{0.0, 0.4, 0.8}
\definecolor{codegray}{rgb}{0.5, 0.5, 0.5}

\lstdefinestyle{cstyle}{
  language=C++,
  keywordstyle=\bfseries\color{codeblue},        
  commentstyle=\itshape\color{codegreen},         
  stringstyle=\color{codeorange},                 
  emphstyle=\color{codepurple},                   
  emph={pickle, config, pickle_kernel, PickleConfig, getPickleDevicePtr, sendConfig, getKernelChannel},  
  numberstyle=\tiny\color{codegray},
  basicstyle=\ttfamily\scriptsize,
  showstringspaces=false,
  tabsize=4,
  breaklines=true,
}

\newcommand{\hl}[1]{\begingroup\setlength{\fboxsep}{0pt}\colorbox{green!15}{#1}\endgroup}

\newenvironment{tightequation}{%
  \begingroup
  \begin{equation*}%
}{%
  \end{equation*}%
  \endgroup
}

\newcommand{\prefetcher}{Pickle}

\newcommand{\todo}[1]{}
\newcommand{\speedup}[1]{$#1\times$}
\newcommand{\data}[1]{#1}

\newcommand*\circled[1]{\tikz[baseline=(char.base)]{
            \node[shape=circle,fill,inner sep=1pt] (char) {\textcolor{white}{#1}};}}

\begin{document}

\raggedbottom

\title{Pickle: Precise, Flexible Cross-Core Last-level Cache Data Prefetching for Irregular Memory Accesses}


\author{Hoa Nguyen}
\email{hn@hnpl.org}
\affiliation{%
  \institution{University of California, Davis}
  \city{Davis}
  \state{CA}
  \country{USA}
}
\author{Pongstorn Maidee}
\email{pongstorn.maidee@amd.com}
\affiliation{%
  \institution{AMD Research \& Advanced Development}
  \city{San Jose}
  \state{CA}
  \country{USA}
}
\author{Jason Lowe-Power}
\email{jlowepower@ucdavis.edu}
\affiliation{%
  \institution{University of California, Davis}
  \city{Davis}
  \state{CA}
  \country{USA}
}
\author{Alireza Kaviani}
\email{alireza.kaviani@amd.com}
\affiliation{%
  \institution{AMD Research \& Advanced Development}
  \city{San Jose}
  \state{CA}
  \country{USA}
}



\begin{abstract}
Graph analytics and sparse scientific workloads are dominated by parallel chains of data-dependent, long-latency memory accesses whose patterns are difficult for hardware to infer yet straightforward to express in software.
Conventional hardware prefetchers attempt to recover this information from address streams alone, but false positives lead to substantial memory traffic overhead.
Software-assisted approaches offer greater flexibility but still consume core limited resources.

We propose \prefetcher{}, a software-defined, hardware-managed last-level cache (LLC) prefetcher that follows the decoupled access/execute philosophy.
\prefetcher{} serves as an independent access engine, fully decoupled from core resources, that executes \emph{prefetch kernels} sliced from the original application to bring data into the shared LLC ahead of demand.
Prefetches are guided by a lightweight \emph{prefetch hint protocol} that lets software express runtime program context, which is fundamentally out of reach for pattern-based hardware.


We evaluate \prefetcher{} using full-system, cycle-level simulation of a cluster of 8 high-performance cores, running all GAP benchmark suite algorithms across nine real-world graphs and irregular-access-dominated scientific applications from the NAS parallel benchmark suite.
Over a no-prefetching baseline, \prefetcher{} achieves \speedup{1.49} geomean speedup with only 2\% DRAM traffic overhead on graph algorithms, and \speedup{1.53} with a 4.5\% memory traffic reduction on NAS scatter/gather kernels.
For reference, the state-of-the-art core-private indirect prefetcher achieves \speedup{1.40} but incurs 43\% DRAM traffic overhead on graph workloads, and \speedup{1.36} at zero traffic overhead on scatter/gather kernels, illustrating the challenge of inferring irregular access patterns without application-level context.
\prefetcher{} also composes transparently with private cache prefetchers: combining it with the state-of-the-art indirect or a simple stride prefetcher yields \speedup{1.65}-\speedup{1.66} and \speedup{1.72}-\speedup{1.84} geomean speedup on graph and NAS scatter/gather workloads, respectively.

\end{abstract}

\maketitle
\section{Introduction}
\label{sec:introduction}

Common irregular workloads, such as graph analytics algorithms and sparse scientific kernels, are largely characterized by parallel chains of data-dependent long-latency memory accesses \cite{basak2019analysis,ainsworth2019software}.
Although graph data structures permit concurrent traversal, their compact, pointer-rich layouts impose data-dependent control flow that serializes memory accesses within each core.
Sparse and unstructured scientific kernels share this bottleneck: their compressed, index-driven layouts produce the same dependent address chains, likewise serializing per-core memory accesses.
Modern microarchitectures address this problem by distributing work across multiple cores, each equipped with deep pipelines, large instruction windows, and reorder buffers (ROBs) to exploit instruction-level parallelism (ILP) and memory-level parallelism (MLP).
Yet across these workloads, MLP within each core remains severely constrained by the data-dependent nature of the access stream.

Modern architectures fundamentally lack an expressive mechanism to capture the intent behind a sequence of memory accesses.
Software prefetch instructions such as x86's \texttt{PREFETCH} \cite{amd_documentation,guide2011intel}, ARM's \texttt{PRFM} \cite{arm_architecture_reference}, and RISC-V's \texttt{prefetch.*} \cite{riscv_zicbop} hint at a single future memory access per instruction, too fine-grained to convey multiple levels of indirection.
Vectorized prefetch instructions, such as \texttt{VGATHERPF} \cite{guide2011intel} and \texttt{PRF*} \cite{arm_architecture_reference}, can express data-level parallelism at coarse granularity, but cannot capture the MLP hidden across multiple levels of indirection or data-dependent control flow inherent to irregular workloads.

The decoupled access/execute (DAE) paradigm \cite{smith1982decoupled} offers a principled approach to this problem: it separates a program's memory access stream from its compute stream, allowing the access stream to run ahead and hide memory latency.
DAE is especially attractive for irregular workloads with abundant MLP.
By decoupling the two streams, the access stream can be \emph{optimized independently} of the compute stream, yielding more hardware-friendly access patterns.

Data prefetchers loosely follow the DAE paradigm.
Conventional hardware data prefetchers for irregular memory accesses (IMA) infer memory access stream from observing the data accesses of the compute stream \cite{yu2015imp,fu2024differential,shi2021hierarchical,duong2024new}, or speculatively execute the compute stream to expose more MLP within a core \cite{mutlu2003runahead,hashemi2016continuous,naithani2020precise,naithani2021vector,naithani2022vector,roelandts2024scalar,naithani2023decoupled}.
Software-assisted prefetching follows the DAE paradigm more directly by separating the memory access stream from the compute stream at the compilation time \cite{kim2002design,guo2025ghost}, or by injecting prefetch instructions into the compute stream \cite{jamilan2022apt,zhang2024rpg2,fu2025magellan,talati2021prodigy}.

These approaches share complementary weaknesses.
Hardware-based approaches can introduce significant memory traffic overhead due to false positives, and they lack the flexibility to adapt to different memory access patterns.
Software-assisted approaches offer more degrees of flexibility but have limited ability to manage the prefetches, as throttling, and timing, while cross-core coordination remain difficult.
Critically, both approaches utilize core resources to facilitate prefetching, which can lead to performance degradation due to resource contention, and lack the ability to efficiently handle complex prefetching logic, such as conditional prefetching.

With these observations, we introduce \textbf{\prefetcher{}}, a software-defined, hardware-managed LLC prefetcher fully decoupled from core resources.
Drawing on the DAE philosophy, \prefetcher{} acts similarly to an independent access engine that executes \emph{prefetch kernels}, compact programs sliced from the original application, to bring data into the shared LLC ahead of demand.
\prefetcher{} resides near the LLC with its own cache and translation unit, consuming no core-side resources.
Because the cores' execution context is not directly accessible, we introduce a lightweight \emph{prefetch hint protocol} that conveys the necessary information to make \emph{precise} and \emph{timely} prefetching decisions.
This information, together with our non-speculative hardware mechanism, reduces speculation in both prefetch generation and handling.
Such speculation in data prefetching is the primary source of the wasted memory bandwidth that bottlenecks large many-core systems \cite{jain2024limoncello}.

This architecture intertwines the flexibility of software prefetching with the memory efficiency of NoC-aware hardware prefetching.
By leveraging the prefetch hint protocol, \prefetcher{} can make informed decisions about \emph{what} to prefetch and \emph{when}, decisions that are fundamentally out of reach for pattern-based hardware prefetchers operating on address streams alone.
\prefetcher{} further employs techniques such as prefetch drops, prefetch delegation, and conditional prefetching to improve timeliness and reduce memory traffic overhead; these techniques would require significant modifications to existing core-private prefetchers.
Furthermore, because \prefetcher{} operates at the LLC and does not alter core-level cache hit/miss patterns, it composes transparently with existing private cache prefetchers and introduces no contention for core-side resources.

Our prefetcher architecture enables non-speculative \emph{conditional prefetching}: it conveys high-level program context at fine-grained granularity, allowing \prefetcher{} to make application-aware decisions about which prefetches to issue.
This capability is especially valuable for graph analytics and sparse scientific workloads, which often involves conditional branches \cite{talati2021prodigy} whose taken rates vary widely on different inputs.
By \emph{non-speculatively} evaluating conditional branches at the prefetch engine, \prefetcher{} avoids the unnecessary memory traffic that other prefetching schemes would generate.

We evaluate Pickle using full-system, cycle-level simulation on a cluster of 8 fully modelled high-performance out-of-order cores and MOESI protocol, running GAP benchmark suite algorithms \cite{beamer2015gap} across nine real-world graphs with distinct characteristics, and irregular-access-dominated scientific applications from the NAS parallel benchmark suite \cite{bailey1991parallel} with large problem sizes.

In this paper, we make the following contributions:
\begin{itemize}
    \item \textbf{Architecture.} \prefetcher{}, a fully decoupled, software-defined hardware-managed LLC prefetcher that enables precise prefetching for IMA patterns without consuming core resources (\S\ref{sec:pickle_prefetcher}).
    \item \textbf{Programming model.} An ISA-agnostic interface in which programmers precisely express indirection chains as prefetch kernels and communicate via prefetch hint protocol (\S\ref{sec:programming_model}).
    \item \textbf{Prefetch backend.} Cross-core request coalescing, and LLC prefetch delegation that localize traffic away from private caches while utilizing previously idle LLC capacity (\S\ref{sec:pickle_backend}, \S\ref{sec:pickle_prefetch_optimizations}).
    \item \textbf{Conditional prefetching.} Case studies on SSSP and BC demonstrating both the benefits and limits of incorporating runtime program state into prefetch decisions (\S\ref{sec:case_study}).
    \item \textbf{Evaluation.} Full-system simulation across 45 algorithm-graph combinations and 3 HPC kernels showing consistent speedups, high prefetch usefulness, and low DRAM traffic overhead (\S\ref{sec:evaluation_discussion}).
\end{itemize}

\section{Background, Motivation, and Related Works}

\subsection{Irregular Memory Accesses}
\begin{figure}[t]
    \centering
    \includegraphics[width=\columnwidth]{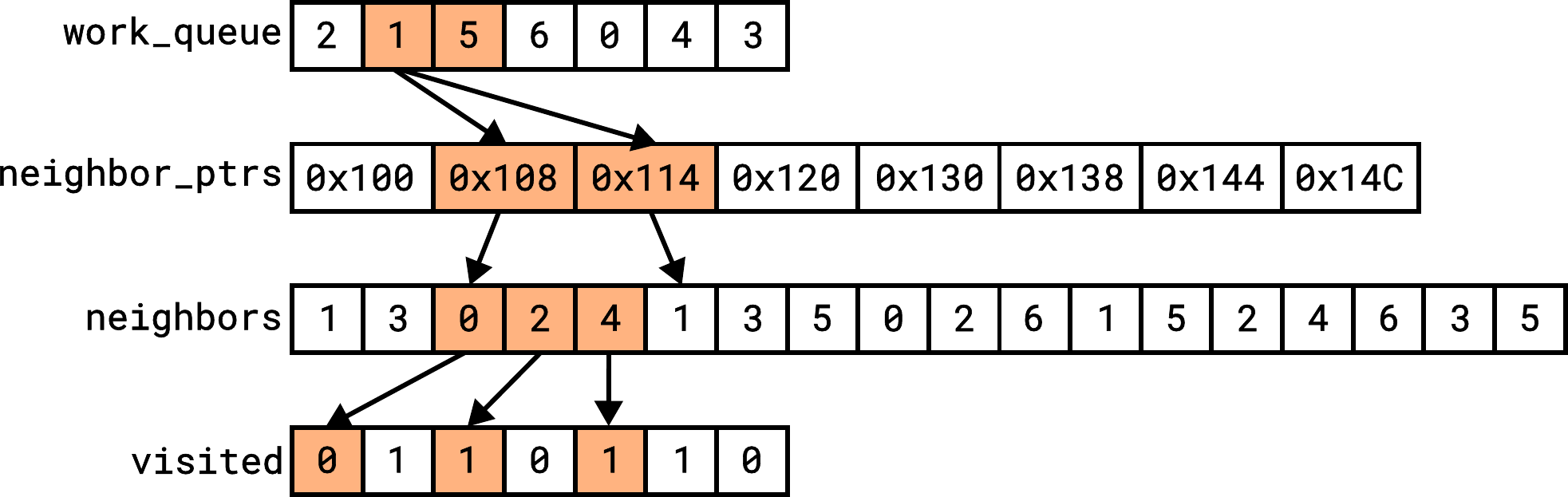}
    \caption{Memory accesses of BFS on a graph in CSR format.}
    \label{fig:graph_example}
\end{figure}

Graph analytics algorithms are widely used in various applications, ranging from social-network processing \cite{beamer2015gap,snapnets} and recommendation systems \cite{eksombatchai2018pixie} to web search \cite{brin1998anatomy}.
As real-world graphs are sparse, they are typically stored in the compressed sparse row (CSR) format (Figure~\ref{fig:graph_example}), which scatters each vertex's neighbor list across memory, inducing IMA patterns.
Scientific workloads exhibit the same behavior through their compressed, index-driven data structures: sparse matrix-vector multiplication (SpMV), the core of conjugate gradient (CG) solvers, stores matrices in CSR and incurs IMA when gathering non-zero elements and their column indices, while unstructured adaptive mesh refinement (UA) incurs IMA when traversing index-encoded mesh connectivity.

These workloads are also rife with data-dependent branches.
In single-source shortest path (SSSP) \cite{beamer2015gap,meyer2003delta,zhang2020optimizing}, a vertex is relaxed only when its tentative distance falls below a threshold; in UA \cite{feng2004unstructured}, an element's contribution is computed conditionally on its spatial location and the refinement criteria.
These branches' outcomes are dependent on runtime-generated values, making the resulting access stream hard for hardware prefetchers to predict.
Additionally, UA's mesh connectivity evolves over time, making it difficult to precisely prefetch without knowledge of the current mesh state.

\begin{figure}[t]
\begin{lstlisting}[style=cstyle, escapechar=@, backgroundcolor = \color{white},frame=none,lineskip=-5pt, basewidth=0.5em,aboveskip=0pt, belowskip=0pt, numbers=left, numberstyle=\tiny\color{gray}, numbersep=5pt]
@\hl{pickle = getPickleDevicePtr();}@
@\hl{config = PickleConfig(distance=32, drop=16,}@
@\hl{\hspace{4em}kernels=libpickle::CreateBFSKernel(work\_queue, g, visited));}@
@\hl{pickle->sendConfig(config);}@
@\hl{pickle\_kernel = pickle->getKernelChannel(0);}@
while (work_queue is not empty) {
    #pragma omp parallel
    {
        local_queue = new Queue();
        #pragma omp parallel for
        for (u_ptr = work_queue.begin(); u_ptr < work_queue.end(); u_ptr++) {
            @\hl{if (u\_ptr + prefetch\_distance < work\_queue.end\_ptr()) \{ }@
            @\hl{\hspace{2em}// send prefetch hint: address of the current node in work\_queue}@
            @\hl{\hspace{2em}*pickle\_kernel = \&(*u\_ptr);}@
            @\hl{\}}@
            for (v in g.neighbors(u)) {
                if (visited[v] == false) {
                    compare_and_swap(visited[v], false, true);
                    local_queue.push(v);
                }
            }
        }
        work_queue.extend(local_queue);
    }
}
\end{lstlisting}
\caption{Top-down implementation of the BFS algorithm. The highlighted lines are the Pickle API calls.}
\label{fig:bfs}
\end{figure}

%
%

\begin{figure}[t]
\begin{lstlisting}[style=cstyle, escapechar=@,backgroundcolor = \color{white},frame=none,lineskip=-5pt, basewidth=0.5em,aboveskip=0pt, belowskip=0pt, numbers=left, numberstyle=\tiny\color{gray}, numbersep=5pt]
const uint64_t vaddr = work_data + software_hint_distance * queue_element_size;
const uint64_t core_vaddr = per_core_context["latest_hint"];

if (context["indirection_level"] == 1) {
    // drop prefetch hint if the core is too close
    if (core_vaddr + prefetch_drop_distance * queue_element_size < vaddr) {
        halt();
    }
    uint64_t aligned_vaddr = (vaddr >> OFFSET_BITS) << OFFSET_BITS;
    device->PrefetchWithVAddr(
        address = aligned_vaddr,
        callback = lambda(array<uint8_t, 64> data) {
            uint64_t offset = (vaddr - aligned_vaddr) / queue_element_size;
            context["visiting_node"] = data[offset:offset+queue_element_size];
        }
    );
}

if (context["indirection_level"] == 2) {
    uint64_t neighbor_ptrs_start_vaddr = descriptor->get_array("neighbor_ptrs").vaddr_start;
    uint64_t target_vaddr1 = \
        neighbor_ptrs_start_vaddr + context["work_queue_value"] * neighbor_ptrs_element_size;
    uint64_t target_vaddr1_aligned = (target_vaddr1 >> 6) << 6;
    device->PrefetchWithVAddr(
        address = target_vaddr1_aligned,
        callback = lambda(array<uint8_t, 64> data) {
            uint64_t offset = (target_vaddr1 - target_vaddr1_aligned) / neighbor_ptrs_element_size;
            context["neighbor_start_ptr"] = data[offset:offset+neighbor_ptrs_element_size];
        }
    );
    uint64_t target_vaddr2 = 
        neighbor_ptrs_start_vaddr + (context["work_queue_value"] + 1) * neighbor_ptrs_element_size;
    uint64_t target_vaddr2_aligned = (target_vaddr2 >> 6) << 6;
    device->PrefetchWithVAddr(
        address = target_vaddr2_aligned,
        callback = lambda(array<uint8_t, 64> data) {
            uint64_t offset = (target_vaddr2 - target_vaddr2_aligned) / neighbor_ptrs_element_size;
            context["neighbor_end_ptr"] = data[offset:offset+neighbor_ptrs_element_size];
        }
    );
}

if (context["indirection_level"] == 3) {
    uint64_t neighbors_start_vaddr = descriptor->get_array("neighbors").vaddr_start;
    uint64_t target_vaddr = 
        neighbors_start_vaddr + context["neighbor_start_ptr"] * neighbor_element_size;
    uint64_t target_vaddr_aligned = (target_vaddr >> 6) << 6;
    device->PrefetchWithVAddr(
        address = target_vaddr_aligned,
        callback = lambda(array<uint8_t, 64> data) {
            uint64_t offset = (target_vaddr - target_vaddr_aligned) / neighbor_element_size;
            context["neighbors"].append(data[offset:offset+neighbor_element_size]);
        }
    );
}

if (context["indirection_level"] == 4) {
    uint64_t visited_start_vaddr = descriptor->get_array("visited").vaddr_start;
    for (uint64_t i = 0; i < context["neighbors"].size(); i++) {
        uint64_t target_vaddr = 
            visited_start_vaddr + context["neighbors"][i] * neighbor_element_size;
        uint64_t target_vaddr_aligned = (target_vaddr >> 6) << 6;
        device->PrefetchWithVAddr(address = target_vaddr_aligned);
    }
}

\end{lstlisting}
\caption{BFS prefetch kernel.}
\label{fig:bfs-kernel}
\end{figure}

\subsection{Pattern-based Prefetching for IMA}
\label{sec:background_prefetchers}
Conventional hardware prefetchers loosely follow the DAE paradigm, \emph{inferring} the memory access stream from patterns in the compute stream.
Temporal prefetchers \cite{joseph1997prefetching,solihin2002using,wenisch2005temporal,wenisch2009practical,jain2013linearizing,bakhshalipour2018domino,wu2019temporal,wu2021practical,ainsworth2024triangel} track correlations between addresses, while spatial prefetchers \cite{smith2006sequential,baer1991effective,chen1995effective,ishii2011access,somogyi2006spatial,michaud2016best,shakerinava2019multi,shevgoor2015efficiently,kim2016path,bakhshalipour2019bingo,pakalapati2020bouquet} exploit spatial locality, detecting structured address patterns or co-accessed blocks.
Both are ineffective on graph and sparse scientific workloads, whose large working sets while exhibiting little spatial locality.
The first layer of indirection, however, is typically structured (e.g., often strided), and this regularity is the basis of instruction-correlated prefetchers.
IMP \cite{yu2015imp} and DMP \cite{fu2024differential} detect it with a stride prefetcher, then correlate one access's returned data with the next access's effective address to infer data-dependent patterns via fixed-function heuristics; an approach prone to noise and false positives (discussed below).
Newer ML-based approaches, Voyager \cite{shi2021hierarchical} and Twilight \cite{duong2024new}, improve accuracy by learning over the stream of LLC misses, though practicality is limited by the need for large, power-hungry models and training data.

Memory access stream ambiguity presents a fundamental challenge for pattern-recognition-based prefetchers.
For example, the access to \texttt{neighbor\_ptrs[u]} shares identical memory access pattern to \texttt{neighbor\_ptrs[u+1]}, where $u$ is the value of \texttt{work\_queue[i]}.
In this case, a sequence of nodes with a similar number of neighbors will result in a sequence of effective addresses with similar memory access patterns, making it difficult to distinguish the cache misses from \texttt{neighbor\_ptrs[u]} and \texttt{neighbor\_ptrs[u+1]}.
As a result, while a pattern-recognition-based prefetcher can correctly detect the indirect memory access chain
\begin{tightequation}
    \texttt{visited[neighbors[neighbor\_ptrs[work\_queue[i]]]]}
\end{tightequation}
it also can incorrectly detect \begin{tightequation}
    \texttt{visited[neighbors[neighbor\_ptrs[work\_queue[i]+1]]]}
\end{tightequation} as an indirect memory access chain.

\subsection{Beyond Pattern Recognition}
Other techniques follow the execute stream more precisely.
Runahead execution \cite{mutlu2003runahead,hashemi2016continuous,naithani2020precise,naithani2021vector,naithani2022vector,roelandts2024scalar,naithani2023decoupled} speculatively executes past a detected long-latency miss to expose more MLP within a core.
Annavaram et al. \cite{annavaram2001data} similarly reconstruct and execute instruction dependencies to generate accesses ahead of time, but remain tightly coupled to the core, relying on its branch predictor to resolve access-stream control flow that is frequently data-dependent and hard to predict for graph analytics and sparse scientific applications.

Software-assisted prefetching follows DAE more directly, using core resources to work explicitly with the access stream.
Ainsworth et al. \cite{ainsworth2019software}, APT-GET \cite{jamilan2022apt}, RPG$^2$ \cite{zhang2024rpg2}, and Magellan \cite{fu2025magellan} detect chains of memory accesses using compiler passes and inject prefetch instructions directly into the compute stream itself.
APT-GET and RPG$^2$ utilize the flexibility of software to adjust prefetch decisions based on runtime program context, while Magellan applies several prefetch strategies to cover different memory access patterns.
These approaches have an overhead of large amount of prefetch instructions in the compute stream.

Hybrid approaches combine both strengths.
Prodigy \cite{talati2021prodigy} has the compiler find indirect-access chains and encode them into a fixed data structure executed by a hardware prefetcher, but the pre-encoded pattern is too rigid for branch-heavy workloads.
ETPP \cite{ainsworth2018event} is more flexible, with the compiler generating a prefetch kernel for a hardware prefetcher, but offers no mechanism for the data-dependent branches common in graph analytics and sparse scientific workloads.
Both also couple tightly to the core, contending with the main program for its resources.

Beyond prefetching, a broader body of work~\cite{horowitz1996informing,carter1999impulse,kuskin1994stanford} has explored programmable memory systems.
T\"ak\=o~\cite{schwedock2022tako} is a more recent example, extending ETPP~\cite{ainsworth2018event} to expose all memory events for programming.
This enables finer-grained control over memory accesses and can potentially improve performance across a wider range of workloads.
These are orthogonal to \prefetcher{}, which precisely prefetches IMA patterns with potentially complex control flow without significantly altering the programming model.

\subsection{Simultaneous Multithreaded (SMT) Helper Threads for IMA}
Helper threads \cite{kim2002design,kim2004physical,luk2001tolerating,zilles2001execution,zhang2007accelerating,kondguli2019bootstrapping,guo2025ghost} execute a compiler-generated access stream on an SMT core alongside the main program.
Unlike hardware prefetchers, a helper thread leverages core resources, such as the instruction cache, branch predictor, and load-store queue, to run an access stream sliced from the main program, achieving high prefetch accuracy.
Recently, Guo et al. \cite{guo2025ghost} show that, by frequently synchronizing the helper thread with the main thread's critical loop, the helper thread maintains a desirable distance ahead of the main thread, yielding prefetches that are both accurate and timely and thus better performance.

However, the accuracy and aggressiveness of helper-thread prefetching are \emph{fundamentally} constrained by the very core resources and their speculative execution on which the helper thread depends, particularly for branch-heavy access streams.
While this approach improves the utilization of otherwise idle core resources, its tight coupling to the core is problematic.
The helper thread relies on the in-core branch predictor to resolve the control flow of the access stream, which is frequently data-dependent and difficult to predict for graph-analytics algorithms.
An overly aggressive predictor emits inaccurate prefetches, whereas an overly conservative one inhibits dynamic loop unrolling and therefore limits prefetch aggressiveness.
Moreover, because the SMT helper thread issues demand requests for the intermediate levels of an indirection chain, these requests can needlessly trigger the L1/L2 data prefetchers \cite{jain2024limoncello}.
As with other core-coupled prefetchers, inaccurate prefetches also create contention between the helper thread and the main thread, degrading main-thread performance.
Finally, an SMT core is a design choice that is not present in every microarchitecture.
Consequently, SMT helper threading is not a general solution for prefetching IMA patterns.

\subsection{High Core-count Systems Have a Bandwidth Problem!}
\label{sec:background_bandwidth}
As the number of cores in a system increases, the aggregate memory bandwidth demand also increases, leading to a bandwidth bottleneck that can limit the performance of memory-intensive workloads.
This problem is amplified by the fact that many-core systems often have a shared memory hierarchy, where multiple cores compete for access to the same memory resources, leading to contention and increased latency.
Jain et al. \cite{jain2024limoncello} show that, at the data center scale, the wasted bandwidth from inaccurate prefetching accumulates and leads to significant performance degradation, and thus the accuracy and timeliness of prefetching is critical for the performance of memory-intensive workloads on many-core systems.

This bandwidth problem motivates \prefetcher{}, whose the design goals are discussed in the next section.

\section{Design Goals}
The design of \prefetcher{} is guided by the following factors:

\textbf{High Prefetch Accuracy and Timeliness.}
We aim to minimize the amount of speculation in prefetch generation by yielding the decision of \emph{what} to prefetch to software.
The hardware mechanism of \prefetcher{} is designed to be \emph{non-speculative}, containing no pattern recognition or branch prediction logic, and instead relying entirely on the software to provide accurate prefetches.
This design choice allows \prefetcher{} to achieve high prefetch accuracy, which is critical for the performance of memory-intensive workloads on many-core systems as discussed in \S\ref{sec:background_bandwidth}.
As timeliness is also critical for the performance of prefetching, \prefetcher{} makes best effort to maintain the timeliness of prefetches by maintaining a certain distance between the prefetch kernel and the core.

\textbf{Flexible Prefetch Logic.}
Flexibility has two facets: the ability to adapt to different memory access patterns, and the ability to implement different prefetch strategies.
Most IMA patterns abstract as $A[f(B[i])]$, with arrays $A$, $B$ and a function $f$ whose form strongly shapes the access pattern.
In graph analytics, $f$ is the identity for breadth-first search (BFS) but a data-dependent branch for single-source shortest path (SSSP), giving distinct patterns;
in unstructured adaptive mesh refinement (UA) \cite{feng2004unstructured}, $f$ is data-dependent and may even evaluate to an illegal index as the mesh is dynamically refined, which can lead to out-of-bounds accesses if unhandled.

These differences demand different prefetch strategies, which is defined as how the prefetcher decides \emph{what}, \emph{when}, and \emph{how} to prefetch.
UA continually refines and coarsens its mesh, altering both connectivity and element count, so the prefetch logic must track the element count and bound-check every prefetch address.

\textbf{Leveraging Program Semantics.}
Program semantics, such as behaviors of conditional branches, provide rich information about the program's memory access patterns.
By incorporating program semantics, we can make more accurate prefetch decisions compared to hardware-only prefetchers.
However, these behaviors are often difficult to capture at the hardware level due to the complexity of out-of-order execution and branch mispredictions.

\textbf{Modular Design and Resource Isolation.}
Completely decoupling the prefetcher from the cores and private caches allows us to avoid using the core's resources while separating the concerns of designing the core complex and designing the prefetcher, allowing each to be optimized independently.
This is in contrast to conventional data prefetchers that are often tightly integrated with the core.
The area and power constraints within the core's pipeline make it difficult to implement complex prefetch logic for irregular memory access patterns \cite{ayers2020classifying} which only benefits a subset of applications.
However, a modular design also means the prefetcher does not have direct access to the core's architectural state.

\textbf{Maintaining Compatibility with Existing Prefetchers.}
Data prefetchers in commercial systems are efficient for many common memory access patterns.
They rely on observing memory access patterns to detect and calibrate prefetcher parameters.
Thus, the design of the new prefetcher should not interfere with such operations of existing prefetchers, allowing them to work in conjunction to further improve performance.

\section{Pickle Prefetcher}
\label{sec:pickle_prefetcher}
\begin{figure}[t]
    \centering
    \includegraphics[width=\columnwidth]{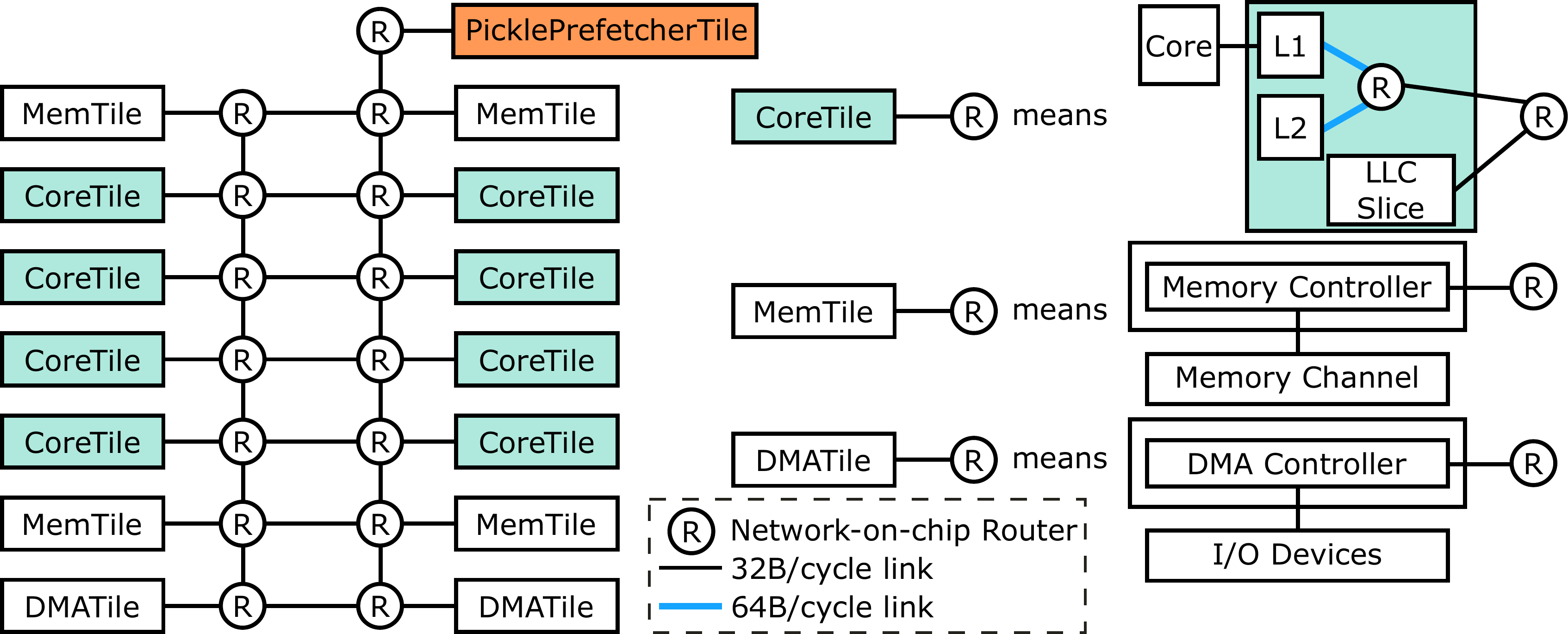}
    \caption{Pickle Prefetcher in the 8-core cluster}
    \label{fig:pickle_in_cache}
\end{figure}

\begin{figure}[t]
    \centering
    \includegraphics[width=\columnwidth]{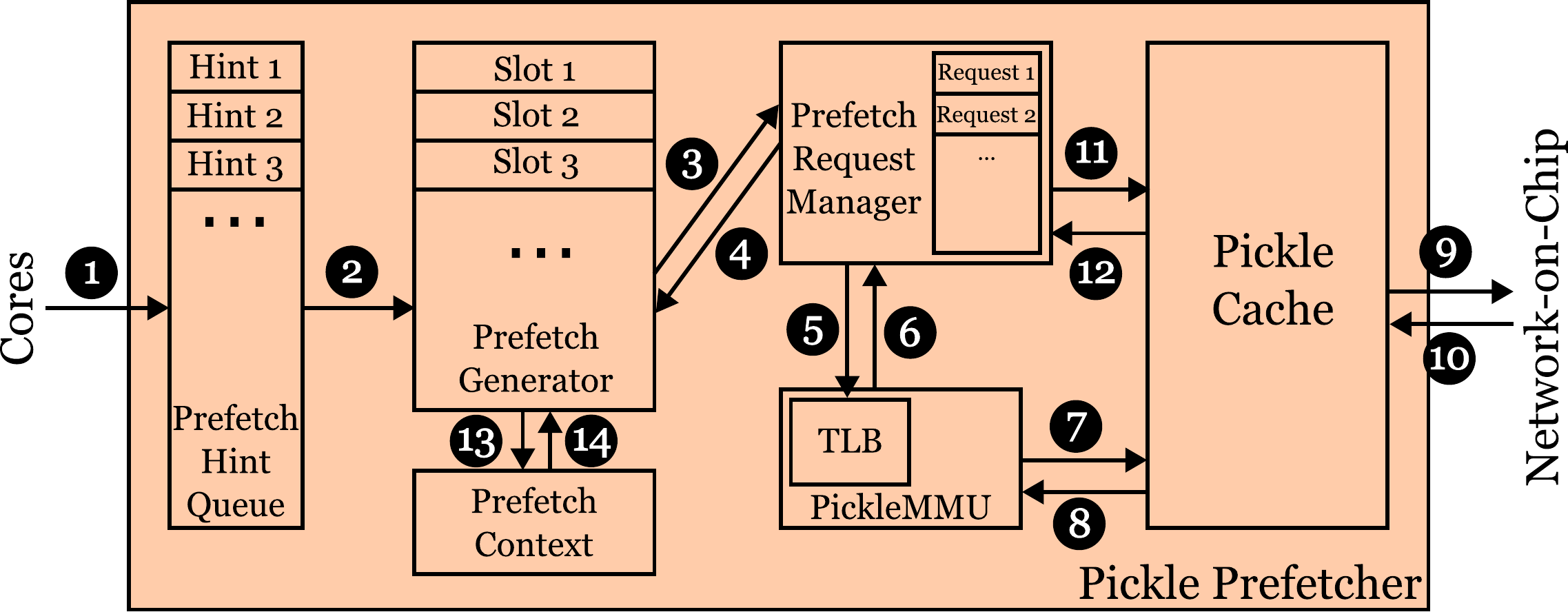}
    \caption{Microarchitecture and data flows of \prefetcher{}}
    \label{fig:pickle}
\end{figure}

\begin{figure}[t]
    \centering
    \includegraphics[width=\columnwidth]{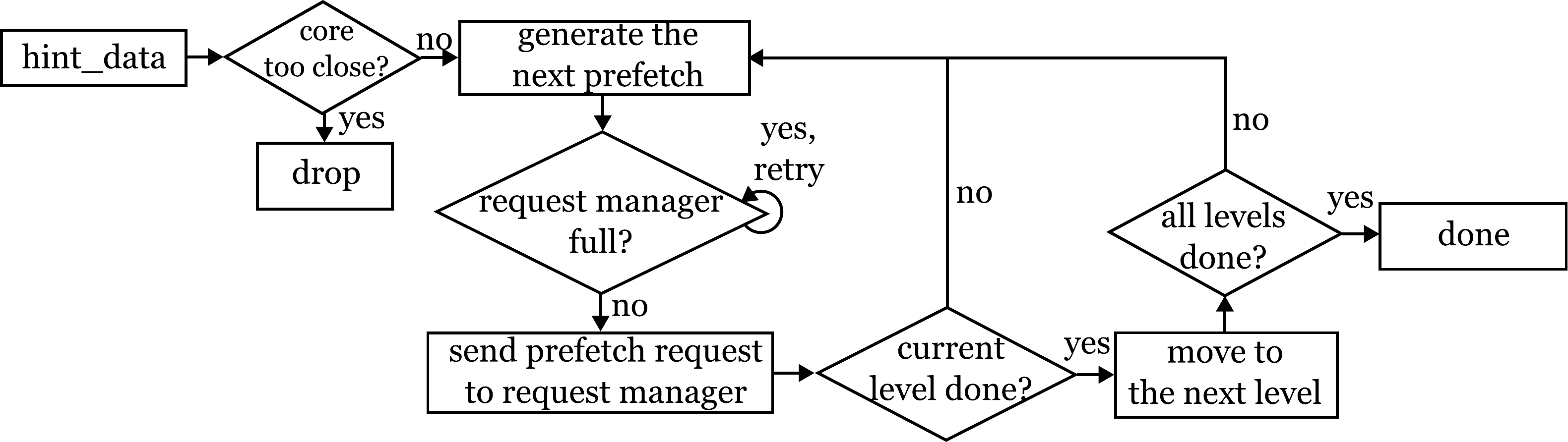}
    \caption{Our implementation of prefetch kernels for graph analytics algorithms}
    \label{fig:pickle_kernel_flow}
\end{figure}

\begin{figure}[t]
    \centering
    \begin{subfigure}[b]{0.40\textwidth}
        \includegraphics[width=\textwidth]{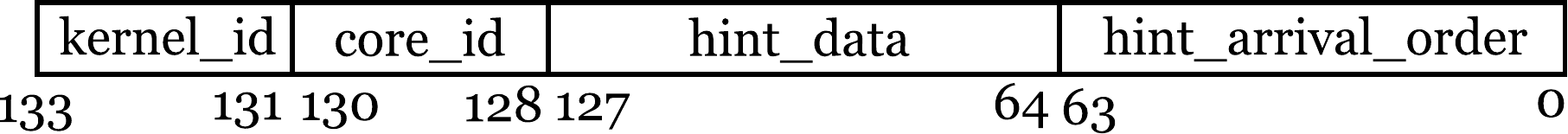}
        \caption{Prefetch hint queue entry}
        \label{fig:data_structure_hint}
    \end{subfigure}
    \hfill
    \begin{subfigure}[b]{0.51\textwidth}
        \vspace{0.5em}
        \includegraphics[width=\textwidth]{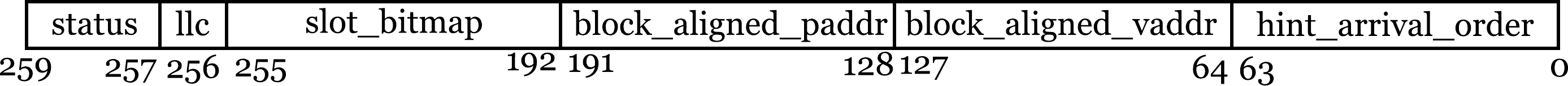}
        \caption{Prefetch request manager entry}
        \label{fig:data_structure_request}
    \end{subfigure}
    \caption{Data structures of \prefetcher{}}
    \label{fig:pickle_prefetcher_implementation}
\end{figure}

Figure~\ref{fig:pickle_in_cache} and Figure~\ref{fig:pickle} illustrates the microarchitecture of \prefetcher{}, which is separated into two parts: the \emph{frontend} and the \emph{backend}.
The frontend, discussed in \S \ref{sec:pickle_frontend}, is responsible for receiving prefetch hints from the cores and generating prefetches.
We provide details of our implementation of prefetch kernels for graph analytics algorithms in \S \ref{sec:pickle_prefetch_kernels}.
The backend, discussed in \S \ref{sec:pickle_backend}, is responsible for managing and issuing the prefetches to the memory system.

Before diving into the details of the frontend, we first discuss how prefetch hints and prefetch kernels are sent to the prefetcher.

\subsection{Programming Model}
\label{sec:programming_model}
The highlighted code in Figure~\ref{fig:bfs} illustrates how prefetching is fully exposed to the programmer at the software level.
\prefetcher{}'s programming model is ISA-agnostic: prefetch hints are regular 64-bit stores to an uncacheable virtual page mapped to the prefetcher's address space, and prefetch kernels are compiled with a separate toolchain and linked into the application binary.
Neither requires ISA extensions nor modifications to the cores.

\subsubsection{Prefetch Kernels}
Prefetch kernels are programmer-supplied routines that each describe one indirect memory access chain of the target algorithm.
They are provided as a library or as part of the application binary and loaded into the prefetch generator engine at the start of execution.
Once triggered, a kernel walks its entire indirection chain, issuing all necessary prefetch requests without further core involvement.
For example, in BFS, a single kernel trigger generates a series of prefetch requests that fetch the \texttt{visited} array entries for all neighbors of the target node.

\subsubsection{Prefetch Hint Protocol}
Prefetch hints are lightweight triggers inserted into the application by the programmer.
Each hint is a 64-bit uncacheable store to the prefetcher's address space, where the store's address selects which kernel to execute and the store's content serves as the kernel's parameter.
For example, if the address range \texttt{0x1000} to \texttt{0x1000 + $N \times 8$} is mapped to the prefetcher, where $N$ is the number of prefetch kernels, then a store to \texttt{0x1000} triggers the first kernel, \texttt{0x1008} the second, and so on.

The overhead of prefetch hints on core performance is negligible.
On X86's architectures, uncacheable stores are strictly ordered among themselves but do not prevent reordering of other memory operations around them \cite{amd_documentation}.
On ARM, stores to Device-nGRE memory are not combined in the store buffers and the core may proceed before the write reaches its endpoint \cite{arm_architecture_reference}.
In practice, each node visit issues only one uncacheable store, limiting the pressure on the cores' load/store queues relative to the graph processing traffic.
On the prefetcher side, a 256-entry hint queue is sufficient for the largest graph in our evaluation, so hints are acknowledged immediately upon arrival.
We fully model uncacheable stores and their interactions with the cores' load/store queues in our evaluation.

In the next subsection, we discuss how the prefetch hints are handled by the prefetcher frontend, and how the prefetch kernels are executed by the prefetch generator engine.

\subsection{Pickle Prefetcher Frontend}

\label{sec:pickle_frontend}

\subsubsection{Prefetch Hint Queue}
Prefetch hints arriving from the cores are buffered in the prefetch hint queue (\circled{1} in Figure~\ref{fig:pickle}) until the prefetch generator engine has a slot available to process them.
Each queue entry (Figure~\ref{fig:data_structure_hint}) contains: a kernel\_id derived from the address of the uncacheable store, identifying which prefetch kernel to execute; a core\_id identifying the sending core; hint\_data, the content of the uncacheable store used as the kernel's input parameter; and hint\_arrival\_order, used later to prioritize prefetches.

\subsubsection{Prefetch Generator Engine and Prefetch Context}
The prefetch generator engine is a cluster of minimal RISC-V cores implementing the RV64E ISA, organized into $K$ slots.
Each slot consists of one core and 1KiB of instruction memory, loaded with prefetch kernels at the start of execution.

When a slot is available, the prefetcher dequeues the front of the hint queue \circled{2} and assigns the hint to that slot.
The slot then uses kernel\_id to select the prefetch kernel to execute from its instruction memory, and the slot only executes one kernel per hint.
The kernel begins executing with hint\_data as its input, generating prefetch requests that are forwarded to the request manager \circled{3}.
As data returns from the memory system \circled{4}, the kernel uses it to walk further levels of indirection, generating additional prefetch requests.

Prefetch context is a scratchpad memory that stores the state needed to drive prefetch decisions, organized at four levels: prefetcher-wide context stores global state shared across all slots; per-core context holds data specific to a given core, such as conditional branch operands (\S\ref{sec:case_study}), shared among all slots prefetching for that core; per-kernel context shared across all slots for the same core and kernel; and per-slot context provides each slot with private working storage during kernel execution.

\subsubsection{Tracking the Core's Progress}
\label{sec:tracking_core_progress}

In addition to triggering a prefetch kernel, the arrival of a prefetch hint from the core also serves to track core's progress.
When a hint arrives, the prefetcher immediately updates the per-kernel context for the corresponding core and kernel with the hint\_data before the hint is assigned to any slot.
This allows all slots serving the same core and kernel to track the core's progress, enabling timely filtering of stale work.
When a hint of the same core and kernel is assigned to a slot, the prefetch kernel checks the per-kernel context to see if the hint is stale.

\subsection{Prefetch Kernels Implementation}
\label{sec:pickle_prefetch_kernels}
For each graph analytics algorithm, we provide a set of prefetch kernels to generate prefetch requests for the algorithm.
Figure~\ref{fig:pickle_kernel_flow} illustrates the high-level flow of our implementation of prefetch kernels, and Figure~\ref{fig:bfs} illustrates our integration of prefetch kernels to the BFS algorithm.
When the prefetcher configuration is sent to the hardware, the software configures the prefetch distance and prefetch drop distance of the prefetch kernels.
Thus, the prefetcher parameters are fully exposed to the software, allowing the software to tune the prefetcher.
Along with the prefetcher parameters, the software also sends the base address and the element size of each data structure that will be accessed by the prefetch kernels.
This information is stored in the prefetch context.

When a prefetch kernel receives the prefetch hint, it first checks if the hint is too late to be useful.
By keeping track of the latest prefetch hint\_data for each core (\S\ref{sec:tracking_core_progress}), the prefetcher can determine if the current hint is too late to be useful.
Note that, across all of our kernel implementations, the prefetch hint\_data is the address of the current node in the work queue when the prefetch hint is sent.
This means software wants to prefetch prefetch\_distance elements ahead of the hint\_data in the work queue.
However, the prefetch hint can be delayed due to the varying execution time of previous prefetch hints.

We drop (i.e., cancel) a prefetch hint when the core has advanced close enough that the prefetch would be unlikely to arrive ahead of demand.
Specifically, we drop a prefetch hint when,
\begin{equation}
    \begin{split}
        \text{hint\_data} &+ \text{prefetch\_distance} \times \text{element\_size} \\
        &\leq \text{latest\_hint\_data} + \text{drop\_distance} \times \text{element\_size}
    \end{split}
    \label{eq:drop_condition}
\end{equation}
where \text{latest\_hint\_data} is the latest prefetch hint\_data for the current core and kernel (\S\ref{sec:tracking_core_progress}), and $\text{hint\_data} + \text{prefetch\_distance} \times \text{element\_size}$ is the address of the element that the software wants to prefetch.
The core is \emph{too close} when the Equation (\ref{eq:drop_condition}) holds.
At the time the prefetch hint is activated, the prefetcher only prefetches when the \emph{effective} prefetch distance is greater than the drop distance.
As shown in \S\ref{sec:prefetch_drop_impact}, this filtering is especially valuable on social graphs, where high-degree nodes cause per-hint execution times to vary widely, allowing the core to overtake the prefetcher.

When a prefetch hint is not dropped, the kernel uses its hint\_data to walk the indirection chain level by level.
In BFS, for instance, the hint\_data encodes the address of a node in the work queue; the kernel fetches the element at $\text{hint\_data} + \text{prefetch\_distance} \times \text{element\_size}$, then uses the returned value to issue prefetch requests for the next level of indirection, continuing until the chain is fully traversed.

Our implementation completes the prefetch requests of one level of indirection before moving to the next level of indirection.
For example, we fetch all neighbors of $u$ before prefetching the \texttt{visited} array of neighbors of $u$.
All fetched neighbors of $u$ are written to the prefetch context, and then used to prefetch the \texttt{visited} array of neighbors of $u$.
This strategy is a departure from the approach of fetching an element of the next level of indirection as soon as the element of the current level of indirection is fetched \cite{yu2015imp,talati2021prodigy,fu2024differential} as it allows us to coalesce the prefetch requests of the same level of indirection.
For each prefetch slot, we initially assign a 1KiB scratchpad memory to store the fetched data, and this can elastically expand at the granularity of 1KiB if the kernel needs more space to store the fetched data.
In our experiment, we use a scratchpad memory with total capacity of 256KiB, and we find that the prefetcher rarely utilizes the entire scratchpad memory capacity as the graphs are typically low-degree on average.

For each prefetch request, the prefetch kernel sends the following information to the request manager: the hint arrival order, block aligned virtual address of the data to be prefetched, the slot ID of the prefetch slot that requests the prefetch, and whether this is a request of the last level of indirection.
The next subsection describes how the request manager uses this information to generate prefetch requests to the memory system.
Further optimization for conditional prefetching is discussed in \S\ref{sec:case_study}.

\subsection{Pickle Prefetcher Backend}
\label{sec:pickle_backend}
The backend of the prefetcher consists of a request manager, PickleMMU, and PickleCache.
At the high level, the frontend sends the virtual address of the data to be prefetched to the request manager, which in turn uses the PickleMMU to translate the virtual address to physical address, and then sends the prefetch requests to the memory system using the PickleCache.

\subsubsection{Prefetch Request Manager}
For each prefetch request sent by the prefetch kernels, the request manager allocates a request entry to store the request information.
Figure~\ref{fig:data_structure_request} illustrates the content each prefetch request generated by the prefetch kernels.

The hint\_arrival\_order is used to determine the priority of the prefetch request, where the hint arrived earlier has higher priority.
When the \prefetcher{}Cache is capable of handling more requests, the request manager sends the prefetch requests with the highest priority first.

The block\_aligned\_vaddr is the block aligned virtual address of the data to be prefetched, and the slot\_bitmap indicates which slots from the prefetch generator request the prefetch to that virtual address.
We coalesce multiple prefetch requests to the same block-aligned virtual address into a single request entry.
When a request is ready to be sent to the memory system, the request manager uses the PickleMMU to translate the virtual address to physical address based on the application's page table \circled{5}\circled{6}.
When a page fault occurs in the PickleMMU, we opt to drop the corresponding prefetch requests.

Finally, the request manager makes decision of sending the prefetch request to the PickleCache \circled{11} or sending the prefetch request to the LLC controller.
If the request is for the last level of indirection, the request manager sends the prefetch request to the LLC controller (discussed in \S \ref{sec:pickle_prefetch_optimizations}); otherwise, the request manager sends the prefetch request to the PickleCache \circled{11}.
When the data from PickleCache is ready \circled{12}, the request manager will update the slot\_bitmap to indicate which slots from the prefetch generator request the prefetch to that virtual address.
The corresponding prefetch kernels will then use the fetched data to generate prefetch requests for the next level of indirection as discussed in \S \ref{sec:pickle_prefetch_kernels}.

Note that, all requests generated by the request manager are treated as demand requests by the cache/memory controllers, so they will be subject to the same priority as the demand requests from the cores.
This is different from conventional prefetching, where prefetch requests are often given lower priority than demand requests.
Additionally, as requests are fetched to either PickleCache or LLC, they do not interfere with the demand requests within the cores' private caches.
This means that the prefetcher does not alter the cache hits/misses patterns in the cores' private caches, making the operations of \prefetcher{} completely \emph{orthogonal} to the observable patterns of the private caches' prefetchers.

\subsubsection{PickleMMU}
The PickleMMU is initialized with the root page table address and relevant information to set up the MMU, which is sent along with the prefetcher configuration during the application initialization.
As the MMU shares the same address space as the cores, any updates to the page table entries must be propagated to the MMU to ensure the correctness of the prefetch requests.
Thus the PickleMMU must participate in TLB shootdowns.

\subsubsection{PickleCache}
The PickleCache is the prefetcher's private cache, and has the same configuration as the L1 data cache of the cores except for larger capacity.
As all prefetch requests \circled{11}\circled{12} and all page walks \circled{7}\circled{8} go through the PickleCache, all data fetched by the prefetcher will be stored in the PickleCache.
As we have the LLC as a victim cache, evicted data from PickleCache will be written to the LLC.
So, by placing fetched data to PickleCache, we effectively prefetch the data to the cache system.

\subsection{Prefetch Optimizations}
\label{sec:pickle_prefetch_optimizations}

\subsubsection{LLC Prefetch Delegation}
\label{sec:llc_prefetch_delegation}

\prefetcher{} integrates into the cache hierarchy as any other requestor would: PickleCache acts as a private cache that participates in coherence, and all prefetch traffic flows through it.
This design requires \emph{no modifications} to the LLC controllers.
However, it forces the final level of indirection, which typically exhibits a scatter-gather pattern whose data the prefetcher never reuses, through the PickleCache, adding congestion.
Figure~\ref{fig:data_source_distribution} shows that this raises PickleCache access latency to \data{133.8} cycles, a \data{127\%} increase from its typical \data{58.9} cycles.

With a minor extension to the LLC controller, \prefetcher{} can instead \emph{delegate} these final-level-of-indirection prefetches directly to the LLC, bypassing the PickleCache entirely.
When the request manager detects that the current request targets the last level of indirection, it translates the address via the PickleMMU and forwards the physical address to the appropriate LLC controller.
As discussed in \S\ref{sec:llc_prefetch_delegation_impact}, this optimization almost universally improves performance by alleviating pressure on the PickleCache.

To implement this optimization, we extend the LLC controllers with a new command, \texttt{FETCH\_IF\_NOT\_PRESENT}, which takes the physical address of the data to be prefetched and an optional timeout parameter.
Upon receiving this command, the LLC controller performs a directory check and tag lookup concurrently: if the block is not present in any caches within the CCD, it issues a fill request to the memory controller; if present in the LLC, it refreshes the block's replacement metadata to most recently used.
The timeout parameter specifies the maximum amount of time between the request arrival to the LLC controller and the request issuance to the memory controller; if the timeout expires, the request is dropped.
While a 10,000-cycle timeout does not result in any timed-out requests in our experiments, we consider it essential for a production deployment where multi-tenant workloads may cause significant variations in LLC access latency.

    \begin{table}
      \footnotesize
     \caption{Default System Parameters Used for Evaluation}
     \resizebox{\columnwidth}{!}{
    \begin{tabular}{ wc{6em} wc{24em}  } 
     \hline
      \textbf{Cores} & \makecell{8 cores @ 4GHz, ARM ISA\\out-of-order; 8-wide pipeline; 448-entry ROB\\64KiB-LTAGE Branch Predictor; 16K-entry BTB; 52-entry RAS}\\ 
      \hline
      \textbf{Core MMU} & \makecell{L1 TLB: 64 entries, fully associative\\ L2 TLB: 1024 entries, 8-way associative\\ 4-level pages with HugePage support, 1 PTW}\\
      \hline
      \textbf{L1 Cache} & \makecell{Capacity: 32KiB for L1I and 48KiB for L1D \\ 8-way set associative for L1I and 12-way set associative for L1D \\ L1I has a stride prefetcher} \\
      \hline
      \textbf{L2 Cache} & \makecell{private unified cache, 1MiB capacity, 16-way set associative} \\
      \hline
      \textbf{L3 Cache} & \makecell{victim cache, mostly exclusive, shared among all cores\\ 32MiB total capacity (4MiB per CoreTile) \\ 16-way set associative} \\
      \hline
      \textbf{Memory} & \makecell{4-channel DDR5 8400; 270GiB/s maximum total bandwidth} \\
      \hline
      \textbf{\makecell{Operating \\ System}} & \makecell{
        Ubuntu Server 24.04.2 LTS Cloud Image\\
        Linux v6.6.71, Transparent Hugepage Enabled
      } \\
      \hline
      \textbf{Guest Compiler} & \makecell{
        gcc 13.3.0\\
      } \\
      \hline
      \textbf{gem5} & \makecell{
        v24.1.0.2
      } \\
      \hline
    \end{tabular}
    }
    \label{table:system}
    \end{table}

\begin{table}
\footnotesize
\centering
\caption{Private Cache Prefetcher and Pickle Configurations}
\begin{tabular}{ | c | c |}
\hline
Prefetchers & Configurations \\
\hline
Stride & \makecell{L1 Stride Prefetcher: Degrees: 4, Table size: 256 entries \\ L2 Stride Prefetcher: Degrees: 16, Table size: 2048 entries.} \\
\hline
DMP & \makecell{L1 Stride Prefetcher: Degrees: 4, Table size: 64 entries \\ L2 Indirect Memory Prefetcher: \\ Index Queue: 8 entries, \\Indirect Candidate Scoreboard: 16 entries, \\ Index Table: 16 entries, Target Table: 16 entries, \\ Range Table: 8 entries, Indirect Relation Table: 16 entries} \\
\hline
Pickle & \makecell{Prefetch Generator: 64 RV64E cores @ 1GHz; 1-stage; in-order, \\ 1KiB instruction memory per core, \\ Prefetch Hint Queue: 256 entries, \\ Prefetch Context: 256KiB Scratchpad Memory, \\ Request Manager: 1024 entries,\\ L1 TLB: 64 entries, fully associative, \\ L2 TLB: 1024 entries, 8-way associative, \\ PickleCache: 256KiB, 16-way set associative \\ LLC delegation enabled with 10,000-cycle timeout \\ SSSP conditional prefetching enabled\\ BC conditional prefetching disabled \\ UA conditional prefetching enabled} \\
\hline
\end{tabular}
\label{table:prefetcher_configuration}
\end{table}

\begin{table}
\footnotesize
\centering
\caption{Graphs Used For Evaluation}
\begin{tabular}{ | c | c | c | c | c |}
    \hline
    \makecell{Graph \\ Names} & \#Nodes & \#Edges & Degrees & \makecell{Active \\ Working \\ Set Size \\ (MiB)}\\
    \hline
    youtube (yt)\cite{yang2012defining} & 1.1M & 3.0M & 2.6 & 28.7\\
    \hline
    web\_google (gl) \cite{leskovec2009community} & 0.9M & 5.1M & 5.8 & 32.8 \\
    \hline
    web\_berkstan (bk) \cite{leskovec2009community} & 0.7M & 7.6M & 11.1 & 39.4 \\
    \hline
    roadNetCA (rd) \cite{leskovec2009community} & 2.0M & 2.8M & 1.4 & 40.6 \\
    \hline
    wiki\_talk (wi) \cite{leskovec2010signed,leskovec2010predicting} & 2.4M & 5.0M & 2.1 & 55.7 \\
    \hline
    as\_skitter (sk) \cite{leskovec2005graphs} & 1.7M & 11.1M & 6.5 & 68.2 \\
    \hline
    pokec (pk) \cite{takac2012data} &1.6M & 30.6M & 18.8 & 141.7 \\
    \hline
    livejournal (lj) \cite{backstrom2006group,leskovec2009community} & 4.8M & 69.0M & 14.2 & 337.2 \\
    \hline
    orkut (or) \cite{yang2012defining} & 3.1M & 117.2M & 38.1 & 493.9 \\
    \hline
\end{tabular}
\label{table:graphs}
\end{table}

\begin{table}
\footnotesize
\centering
\caption{Problem Size of NPB workloads Used For Evaluation}
\begin{tabular}{ | c | c | c | c | c |}
    \hline
    Application & Problem Size & \makecell{Active \\Working \\Set Size} & \makecell{Sampled Kernel} & \makecell{Sample Size} \\
    \hline
    CG & Class E & approx. 150GiB & \texttt{SpMV} & 9 \\
    \hline
    IS & Class D & approx. 16GiB & \texttt{rank()} & 25 \\
    \hline
    UA & Class D & approx. 8GiB & \texttt{transf()} & 30 \\
    \hline
\end{tabular}
\label{table:npb_inputs}
\end{table}

\section{Methodology}

We evaluate \prefetcher{} on a cluster of 8 high-performance cores and 32MiB of LLC using full-system simulation.
We use the gem5-based \cite{binkert2011gem5,lowepower2020gem5} software/hardware codesign framework, Choreographer \cite{nguyen2025choreographer}, which models near-cache accelerators within a fully modeled NoC.
We use the gem5's implementation of ARM AMBA CHI \cite{amba_chi} to model MOESI-like cache coherence protocol.

\subsection{System Overview}
\label{sec:system_overview}
The system configuration used for the experiments is specified in Table \ref{table:system}.
The system with \prefetcher{} is illustrated in Figure~\ref{fig:pickle_in_cache}, and the baseline system is identical to the system with \prefetcher{} except for the absence of the PickleTile.
All the tiles form a network-on-chip (NoC) with a mesh topology.

\subsubsection{NoC Configuration}
\label{sec:noc_configuration}
The NoC is modeled using gem5's RUBY implementation of the CHI protocol \cite{amba_chi}, which is used in ARM platforms.
We configure the CHI cache to a MOESI-like protocol where the shared dirty state is allowed.
The L3 cache is a victim cache, and mostly exclusive of the L1 and L2 caches.
The L1 cache, L2 cache, and each slice of the L3 cache are configured to handle 16, 64, and 256 concurrent requests, respectively.
The \prefetcher{}Cache is configured to handle 64 concurrent requests.

%
\subsection{Baseline Systems}
\label{sec:baseline_systems}
Table \ref{table:system} provides the details of all components we use for modeling the system.
We compare \prefetcher{} with baseline systems with different private cache data prefetchers,
\begin{itemize}[leftmargin=1em]
    \item \textbf{Stride prefetcher}: A system with an L1 stride and an L2 stride prefetcher, both operate entirely on physical address space.
    \item \textbf{Differential matching prefetcher (DMP)}: A system with DMP, the state-of-the-art (SOTA) data prefetcher for indirect memory accesses \cite{fu2024differential}.
    DMP consists of an L1 stride prefetcher that prefetches to L1 cache, and an L2 prefetcher prefetching indirect data accesses to L2 cache.
    The L1 prefetcher operates in physical address space, while the L2 prefetcher operates in virtual address space and uses the core's MMU for translation.
    We model a 7-cycle latency for communication between the core MMU/L1 cache and the L2 prefetcher.
    As the link to the DMP implementation cited by \cite{khadem2025dx100} is unavailable, we faithfully implement DMP and use the evaluation parameters from the DMP paper \cite{fu2024differential}.
\end{itemize}

Different from the default behavior of gem5's ARM MMU, which does not perform page table walking for prefetch requests, we implement page table walking for prefetch requests.
This change consistently improves the performance of systems involving DMP.


\subsection{Workloads}
\label{sec:workloads}

\subsubsection{Graph Analytics Algorithms}
We evaluate \prefetcher{} on multithreaded GAP benchmark suite \cite{beamer2015gap} implementations of betweenness centrality (BC), breadth-first search (BFS), connected components (CC), pagerank (PR), and single-source shortest paths (SSSP).
We exclude triangle counting, which requires undirected graphs: only 3 of our 9 graphs qualify and complete within our 14-day budget, which is too few for reaching meaningful conclusions.

These 5 algorithms across 9 graphs (Table~\ref{table:graphs}) and 5 prefetcher configurations (Table~\ref{table:prefetcher_configuration}) yield over 200 experiment points.
Figure~\ref{fig:overall_performance} presents all data; subsequent analysis focuses on geometric means and representative algorithm subsets.

A small number of configuration points are missing: SSSP-roadNetCA lacks a baseline due to a gem5 out-of-order core mis-executing Arm's STP instruction, and DMP times out on BC-orkut, CC-roadNetCA, and PR-wiki\_talk from excessive prefetch volume.
\prefetcher{}+DMP similarly times out on CC-orkut and PR-web\_google; we expect its performance to track the \prefetcher{}-only configuration on these pairs.
These data points are not included in the performance analysis.

\subsubsection{Sparse Scientific Kernels}
Table~\ref{table:npb_inputs} evaluates \prefetcher{} on large-input sparse scientific kernels from the NAS Parallel Benchmark suite \cite{bailey1991parallel}.
We sample each kernel's scatter/gather hot spot, its most memory-intensive phase: the radix-sort scatter phase of integer sort (IS), the sparse matrix-vector multiply (SpMV) phase of conjugate gradient (CG), and the heat-transfer forward phase of unstructured adaptive mesh refinement (UA).
For CG and UA, we inject \prefetcher{} hints directly into the legacy Fortran code base by interfacing our C++ prefetching library to it, showing \prefetcher{} can accelerate legacy applications without extensive code changes, highlighting the flexibility of our programming model.

\subsubsection{Prefetch Kernel Generation}
As presented so far, \prefetcher{} requires programmer expertise to identify where to place prefetch requests, slice out the prefetch code, and write the prefetch kernels.
All kernels are written manually (\S\ref{sec:pickle_prefetch_kernels}) except those for IS and CG, which we found LLMs can produce effectively.
Given examples from GAPBS \cite{beamer2015gap}, the Claude Opus 4.6 \cite{claude2026} coding agent injects prefetch hints and produces correct kernels for NPB's IS and CG without manual intervention.
For UA, the LLM correctly identifies the scatter/gather operations in the heat-transfer phases, but its generated kernel is incorrect and required manual fixing.

\subsubsection{Measurement}
For each algorithm-input-prefetcher configuration, we run two phases: a warm-up phase that populates the caches and lets private-cache prefetchers learn the access patterns, and a measurement phase that collects performance metrics.
Checkpoints are generated before the warm-up phase.

\section{Evaluation \& Discussion}
\label{sec:evaluation_discussion}

By default, we use the \prefetcher{} configuration shown in Table \ref{table:prefetcher_configuration} unless otherwise stated.
All results are multi-core results, where the benchmark runs on all 8 cores with 8 software threads managed by OpenMP.
During the runtime, the graph data structures are stored in 4KiB pages.
Therefore, TLBs of all cores do not cover the entire data structures for all inputs and all applications.

\subsection{Performance Analysis}
\label{sec:performance_analysis}
We focus primarily on the impact of each prefetcher on overall performance and memory-system utilization.
We use additional DRAM traffic relative to the no-prefetching baseline as a proxy for prefetch volume, since multiple prefetch requests may be coalesced into a single DRAM access.
When computing geometric means, we include only benchmarks for which all prefetcher configurations completed within the 14-day timeout.
Although TC data is available, it's only applicable on too few graphs in our experiment to draw meaningful conclusions.
When we refer to speedups, we are referring to the speedup of the prefetcher configuration over the non-prefetching baseline unless otherwise specified.

\subsubsection{Overall Performance}
\label{sec:overall_performance}

\begin{figure}[t]
    \centering
    \begin{subfigure}[b]{\columnwidth}
        \centering
        \includegraphics[width=\columnwidth]{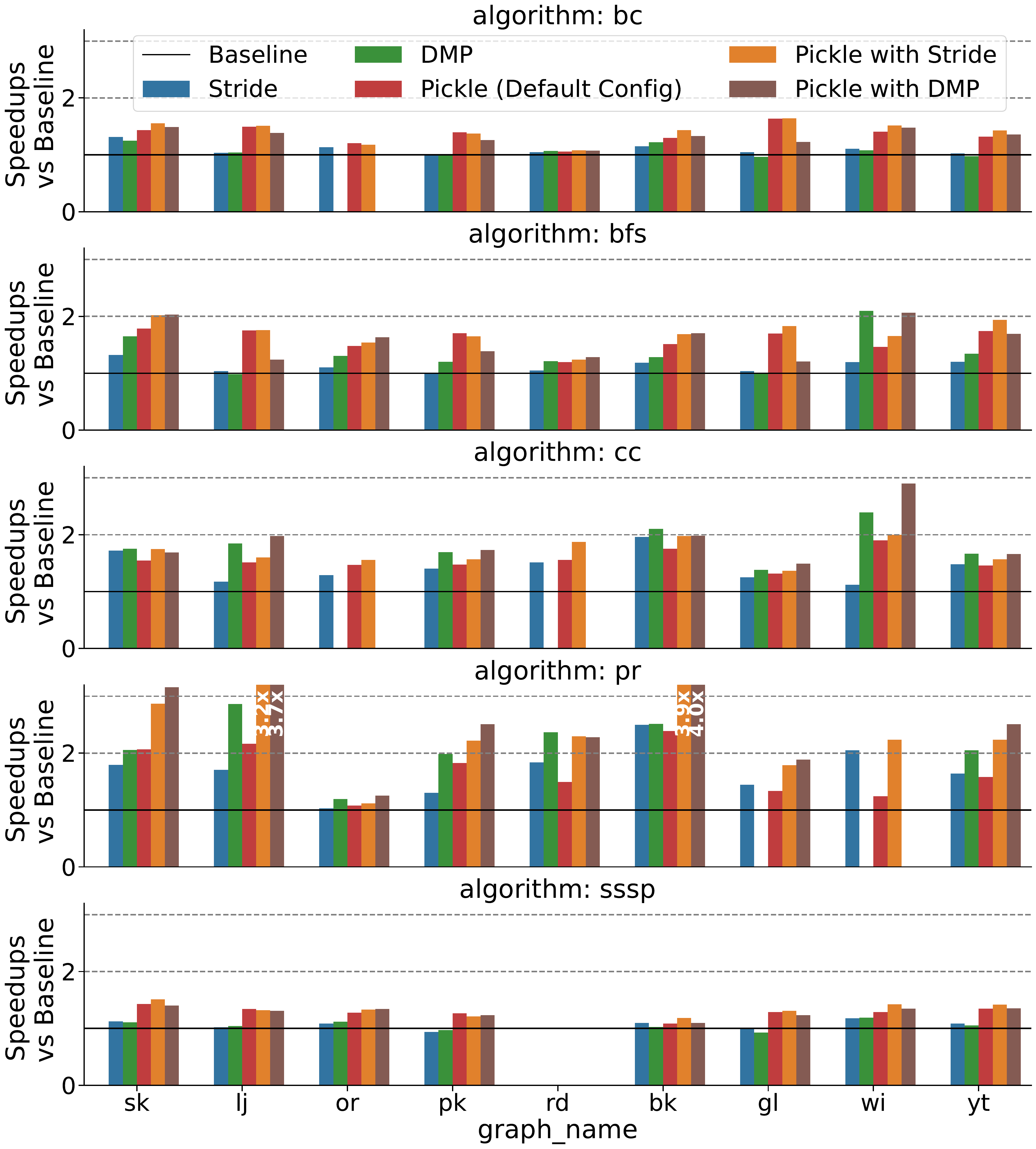}
        \caption{GAP benchmark suite}
        \label{fig:overall_performance_a}
    \end{subfigure}
    
    \begin{subfigure}[b]{\columnwidth}
        \centering
        \includegraphics[width=\columnwidth]{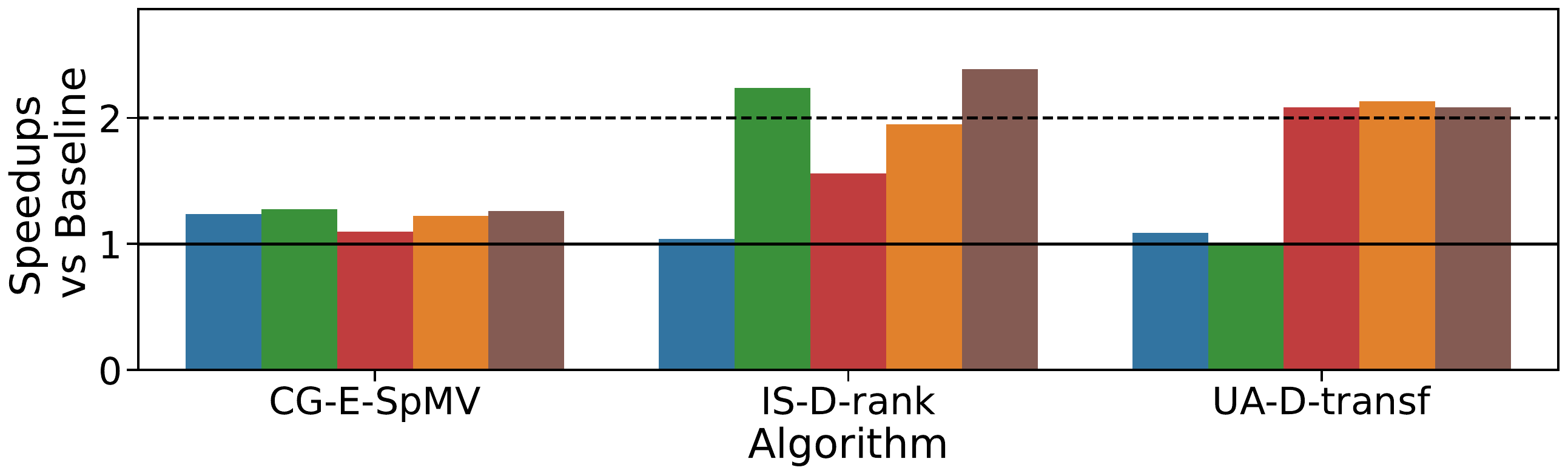}
        \caption{NAS benchmark suite}
        \label{fig:overall_performance_b}
    \end{subfigure}
    \caption{Performance of all benchmarks under each prefetcher configuration discussed in \S \ref{sec:overall_performance}. Missing bars indicate simulations that exceeded the 14-day timeout.}
    \label{fig:overall_performance}
\end{figure}

Figure~\ref{fig:overall_performance} shows the speedups of different prefetcher configurations over the non-prefetching baseline.
The data reflects two key trends: \prefetcher{} outperforms private-cache prefetchers on average, and combining \prefetcher{} with private-cache prefetchers yields additional speedups.
\S \ref{sec:prefetcher_perspective} and \S \ref{sec:memory_system_perspective} reveal the contrast in how these prefetchers achieve these results.

\paragraph{Pickle vs.\ private-cache prefetchers.}
Across graph workloads, \prefetcher{} yields a geomean speedup of \speedup{\data{1.49}} versus \speedup{\data{1.40}} for DMP; on the NAS benchmarks it reaches \speedup{\data{1.53}} versus \speedup{\data{1.36}}.
However, relative performance varies by algorithm.
Where performance-critical indirection chains are under data-dependent branches (BC, SSSP, UA) or are subject to noises (BC, BFS, SSSP), as discussed in \S\ref{sec:background_prefetchers}, \prefetcher{} achieves better performance with geomean speedups of \speedup{\data{1.39}} (BC), \speedup{\data{1.59}} (BFS), \speedup{\data{1.28}} (SSSP), and \speedup{\data{2.08}} (UA) against DMP's \speedup{\data{1.08}}, \speedup{\data{1.35}}, \speedup{\data{1.04}}, and \speedup{\data{1.01}}.
Where the chain is branch-free and base arrays are traversed sequentially (CC, PR, CG, IS), DMP leads, at \speedup{\data{1.73}}, \speedup{\data{2.04}}, \speedup{\data{1.28}}, and \speedup{\data{2.24}} versus \prefetcher{}'s \speedup{\data{1.55}}, \speedup{\data{1.69}}, \speedup{\data{1.10}}, and \speedup{\data{1.56}}.
With the prefetch usefulness data in \S\ref{sec:prefetcher_perspective}, this shows that when the access pattern is largely free of noise and control flow, DMP's pattern detection is accurate enough to prefetch useful data into the private caches; \prefetcher{}, being immune to such noise and control flow, instead leans on its high accuracy to achieve better performance than DMP despite prefetching farther from the core.

\paragraph{Composition with private-cache prefetchers.}
\prefetcher{}'s fully decoupled design composes naturally with existing private-cache prefetchers.
Paired with DMP, it reaches \speedup{\data{1.65}} on graph algorithms and \speedup{\data{1.84}} on NAS (vs.\ \speedup{\data{1.40}} and \speedup{\data{1.43}} for DMP alone); paired with a stride prefetcher, \speedup{\data{1.66}} and \speedup{\data{1.72}} (vs.\ \speedup{\data{1.23}} and \speedup{\data{1.12}}).
This reflects \prefetcher{}'s modular design, which lets each prefetcher specialize: stride prefetchers capture sequential accesses (e.g., \texttt{memcpy()}) and pull data close to the core, while \prefetcher{} either handles the indirection chains DMP cannot detect or, for the chains DMP can detect, prefetches further ahead than DMP.
This suggests \prefetcher{} can also compose with related techniques, such as helper-thread prefetchers, as long as the two target different prefetch distances.

\subsubsection{The Core's Perspective: Load-to-use Latency, The Shift in Data Movement, \& Traffic to Core's MMU}
\label{sec:core_perspective}

\paragraph{Impact on Load-to-use latency.}
For graph algorithms, stride prefetcher, DMP, and \prefetcher{} achieve geomean reductions of load-to-use latency by a factor of \data{40}\%, \data{44}\%, and \data{49}\%, respectively.
For NPB kernels, as for each sample point, each prefetcher configuration progresses differently in the sampled period, we are unable to directly compare the reductions in load-to-use latency directly.

Despite placing data at the LLC rather than the private caches, \prefetcher{} converts a significant fraction of long-latency DRAM accesses to LLC hits, which has an outsized effect on average load-to-use latency.
Stride prefetchers, by contrast, primarily accelerate sequential accesses that are rarely on the critical path, and DMP's pattern-based detection produces inaccurate prefetches.

\paragraph{The Shift in Data Movement.}
\begin{figure}[t]
    \centering
    \includegraphics[width=\columnwidth]{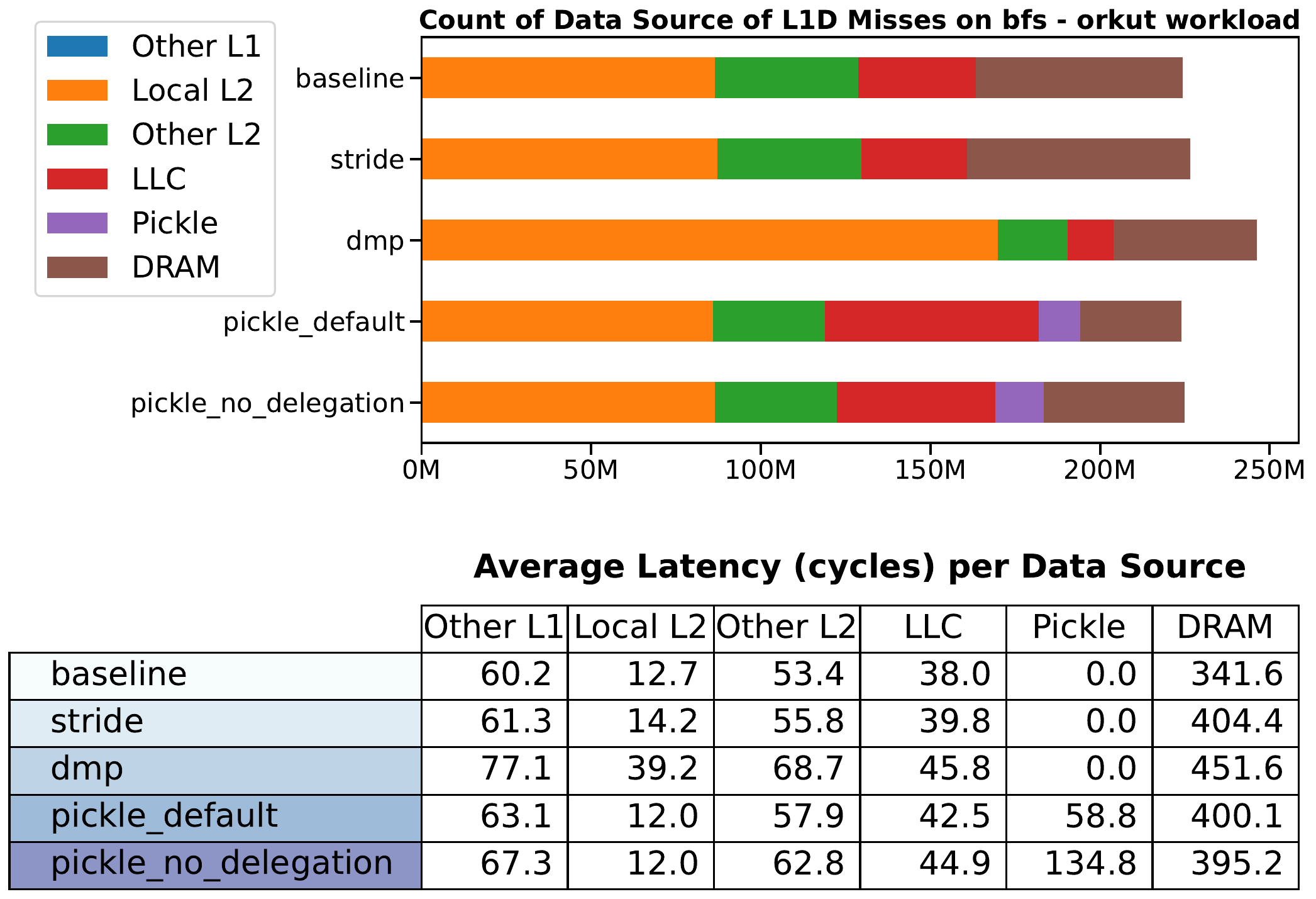}
    \caption{Location and average access latency per location of data being accessed after L1D demand misses for BFS on the orkut graph.}
    \label{fig:data_source_distribution}
\end{figure}
Figure~\ref{fig:data_source_distribution} shows that both \prefetcher{} and DMP reduce the number of demand L1D misses serviced by DRAM, but through different mechanisms.
Compared to the baseline, \prefetcher{} shifts a portion of DRAM access to LLC and \prefetcher{}Cache accesses while slightly increase the average access latency of LLC from \data{38.0} cycles to \data{42.5} cycles.
In contrast, DMP significantly increases the fraction of L1D misses serviced by the L2 cache, but at the cost of inflating average L2 access latency by a factor of \data{3.1}.
This increased L2 pressure stems from DMP prefetching useful data into L2 while simultaneously polluting L1D with unnecessary data.
The stride prefetcher configuration confirms that this L1D pollution is attributable to DMP itself rather than its L1 stride prefetcher.

\paragraph{Pressure on Cores' MMUs}

\begin{figure}[t]
    \centering
    \includegraphics[width=\columnwidth]{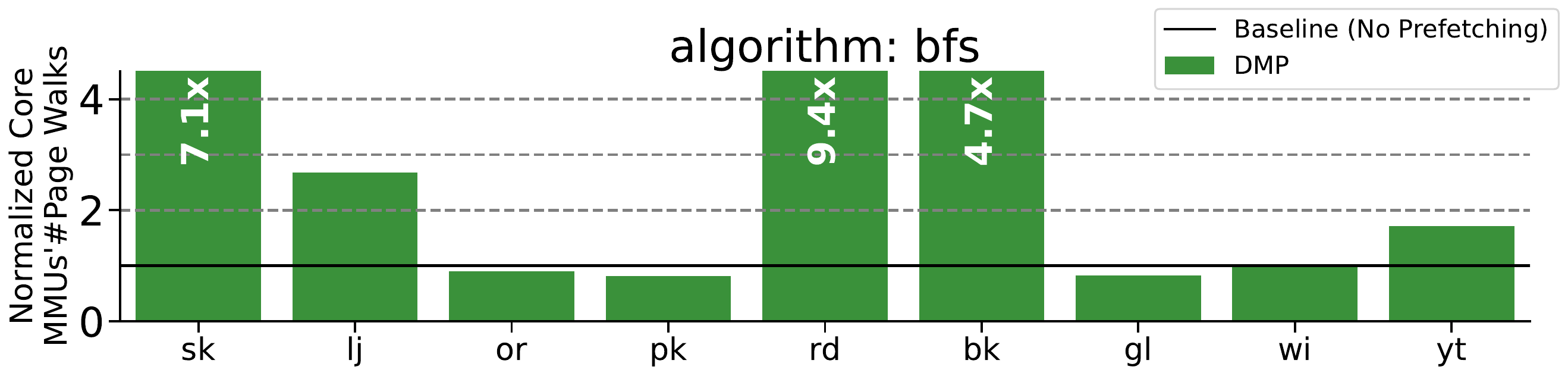}
    \caption{Number of page walks handled by cores' MMUs normalized to non-prefetching baseline.}
    \label{fig:core_mmu_walks}
\end{figure}
Figure~\ref{fig:core_mmu_walks} shows the significant increase in the number of page walks handled by cores' MMUs for DMP on BFS.
This suggests that DMP's L2 indirection prefetcher's false detection puts pressure on the cores' MMUs to translate a large amount of addresses, contributing to the observed L1D pollution.
In contrast, \prefetcher{} uses its dedicated MMU, avoiding any additional page walk overhead on the cores.

\subsubsection{The Prefetcher's Perspective: Prefetch Usefulness}
\label{sec:prefetcher_perspective}

\begin{figure}[t]
    \centering
    \includegraphics[width=\columnwidth]{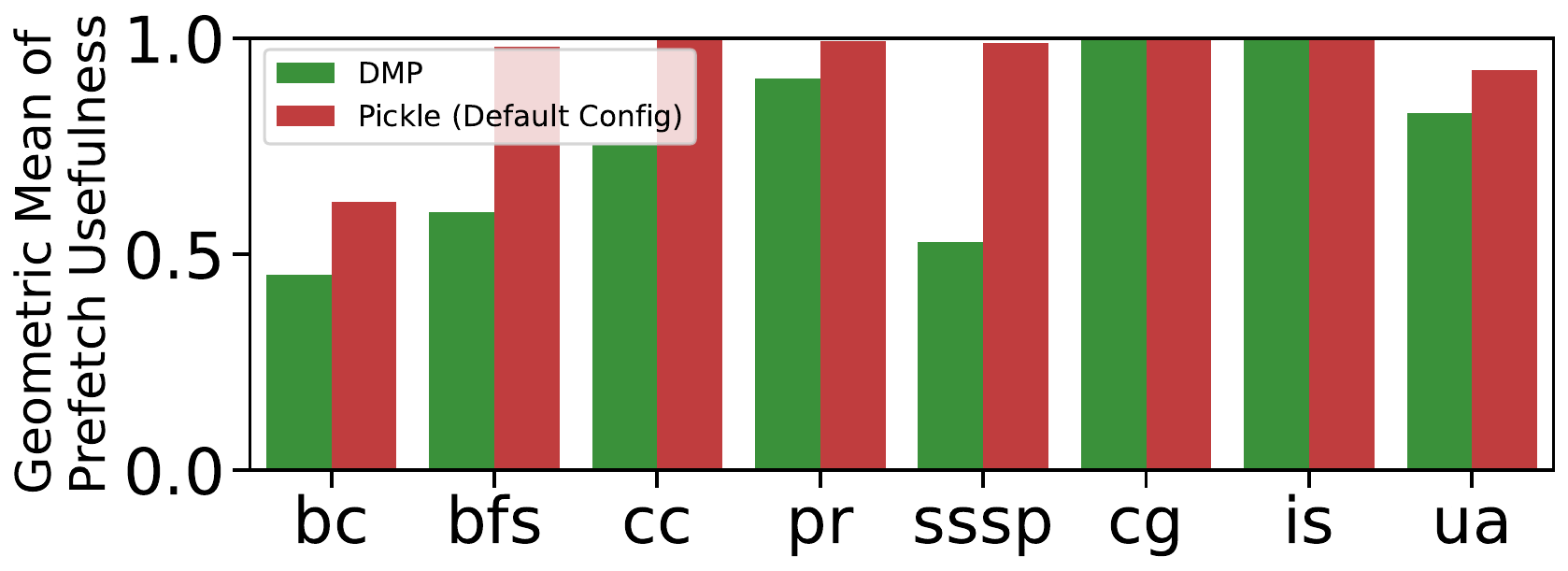}
    \caption{Geometric mean of prefetch usefulness.}
    \label{fig:geomean_prefetch_usefulness}
\end{figure}

Figure~\ref{fig:geomean_prefetch_usefulness} shows that \prefetcher{} achieves near-perfect usefulness across most applications except BC, whereas DMP's prefetch usefulness varies significantly.
A prefetch is considered useful if the prefetched cache line is brought into any level of the cache hierarchy and consumed by the core before being evicted.
These results show that, although DMP can be highly accurate, it is prone to noise that limits its effectiveness.
Because \prefetcher{} handles conditional branches non-speculatively, it achieves high prefetch accuracy, as reflected in the prefetch usefulness of SSSP and UA.
For BC, however, we trade prefetch accuracy for more performance, as discussed in more detail in \S\ref{sec:case_study}.

While prefetch usefulness captures the desired interaction between demand requests and the prefetcher, it does not reflect a program's overall performance.
For example, even though DMP attains high prefetch usefulness, it misses most of the important indirection chains in UA, issuing 98.97\% fewer prefetches than \prefetcher{} on that workload.
This directly translates to the low performance improvement shown in \S\ref{sec:overall_performance}.

\subsubsection{The Memory System's Perspective: Memory Traffic Overhead}
\label{sec:memory_system_perspective}

\begin{figure}[t]
    \centering
    \includegraphics[width=\columnwidth]{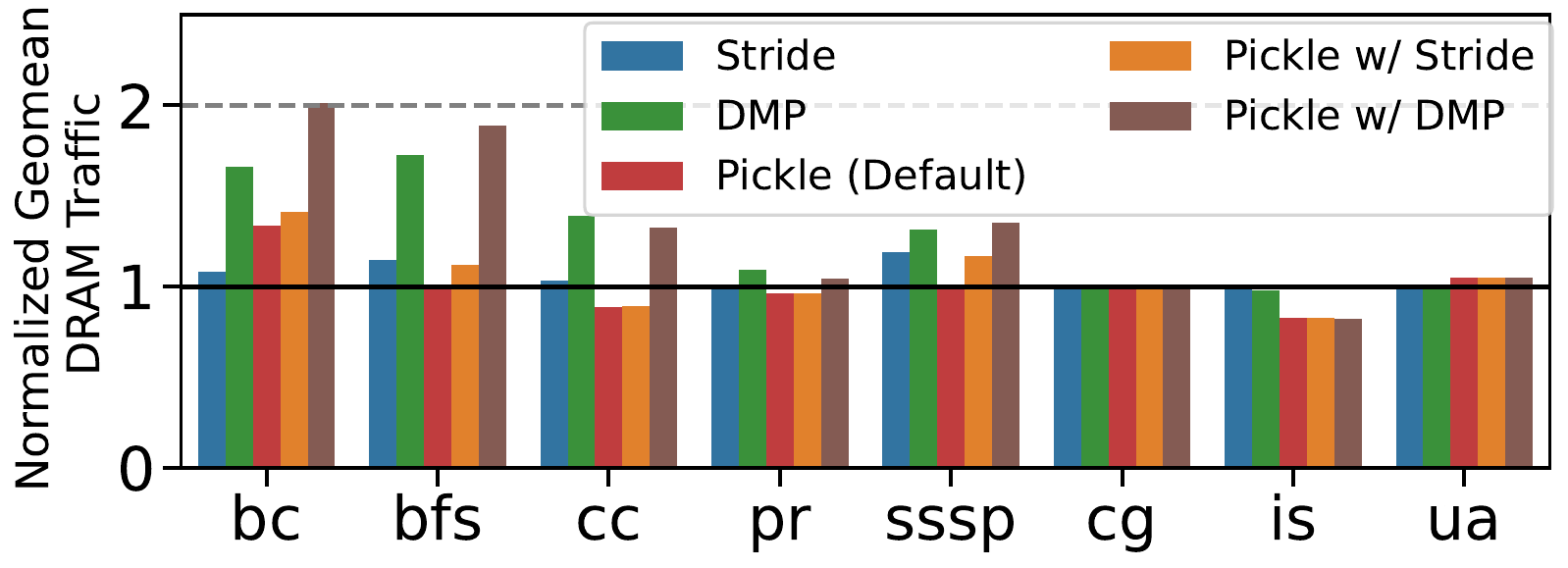}
    \caption{Memory overhead of various prefetchers.}
    \label{fig:memory_overhead}
\end{figure}

Figure~\ref{fig:memory_overhead} shows that, for the graph workloads, the accuracy of \prefetcher{}'s prefetches yields a geometric mean memory traffic overhead of just \data{2\%}, compared to \data{9\%} for the stride prefetcher and \data{43\%} for DMP.
For the NPB workloads, the \prefetcher{} reduces the memory traffic per processed element by 4.5\% as compared to the baseline, while DMP and stride maintain no traffic overhead over the baseline.
We hypothesize that the reduction of memory traffic comes from two primary sources: reducing execution time of the program reduce the memory traffic from background processes, and our scheme of marking data in LLC (\S\ref{sec:llc_prefetch_delegation}) as most recently used keep useful data from being evicted.
Notably, BC suffers from over-prefetching due to tradeoffs in prefetch accuracy and performance, whereas SSSP benefits from branch-aware optimization and achieves near-baseline memory traffic overhead.

\begin{figure}[t]
    \centering
    \includegraphics[width=\columnwidth]{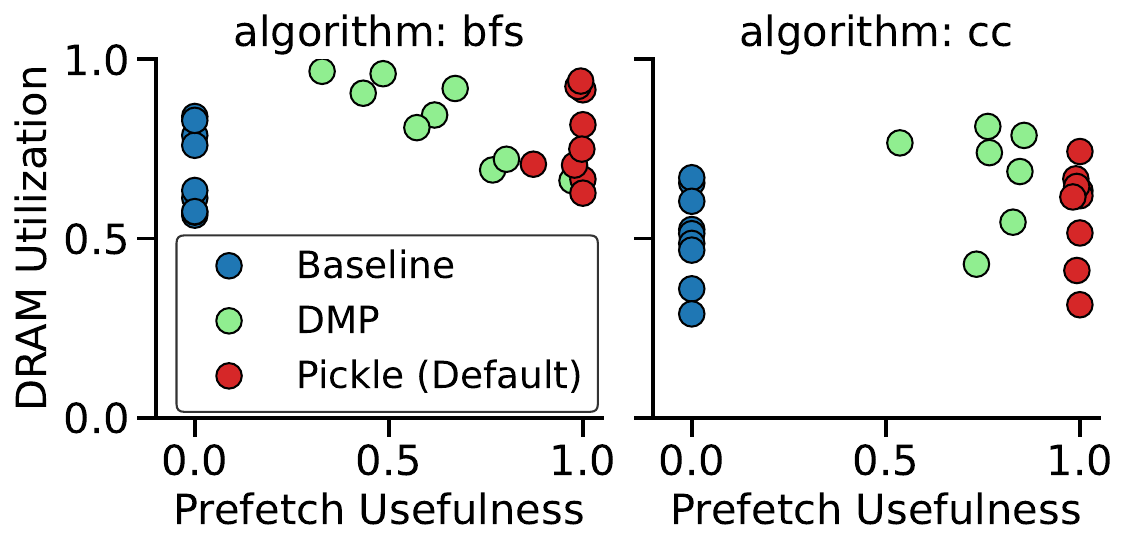}
    \caption{DRAM utilization versus prefetch usefulness.}
    \label{fig:dram_utilization_vs_prefetch_usefulness}
\end{figure}

Figure~\ref{fig:dram_utilization_vs_prefetch_usefulness} plots DRAM utilization against prefetch usefulness; an ideal prefetcher would occupy the lower-right quadrant, achieving high usefulness with minimal additional DRAM utilization over the baseline.
We define DRAM utilization as the fraction of time the DRAM is servicing at least one request.
Both \prefetcher{} and DMP increase DRAM utilization relative to the baseline, but a larger fraction of \prefetcher{}'s additional DRAM activity serves useful prefetches.
This is consistent with the memory traffic overhead in Figure~\ref{fig:memory_overhead}, and shows that DRAM utilization remains unsaturated for several workloads, leaving headroom for additional prefetching.

\subsection{Sensitivity Analysis}
\subsubsection{Impact of LLC Capacity}

\begin{figure}[t]
    \centering
    \includegraphics[width=\columnwidth]{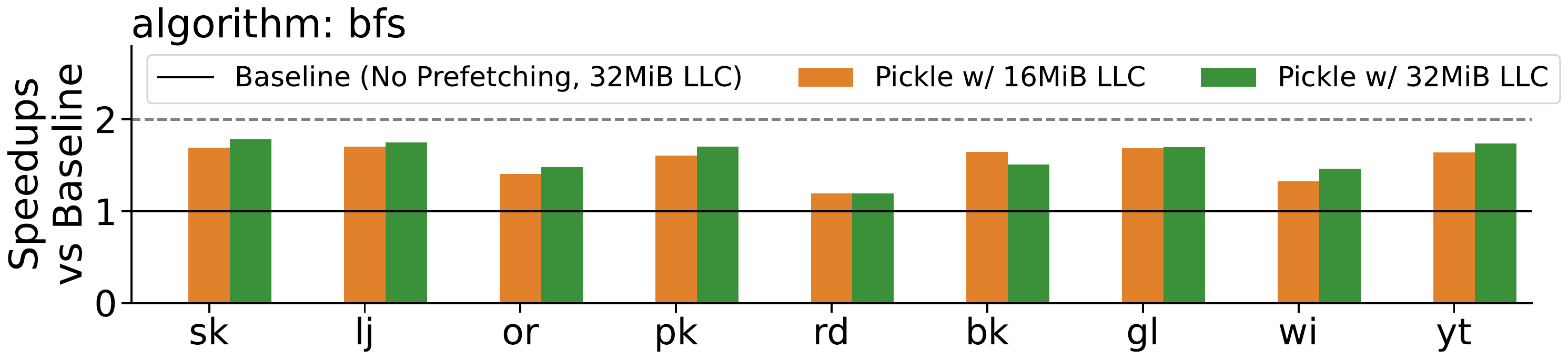}
    \caption{Speedup of \prefetcher{} with 16MiB LLC and 32MiB LLC over non-prefetching baseline with 32MiB LLC.}
    \label{fig:pickle_llc_size_sensitivity_speedups}
\end{figure}

Figure~\ref{fig:pickle_llc_size_sensitivity_speedups} illustrates the trade-off between increasing LLC capacity and deploying \prefetcher{} with a smaller LLC.
As shown in \S\ref{sec:area_power_overhead}, \prefetcher{}'s area overhead is less than that of an additional 16MiB of LLC, yet \prefetcher{} with a 16MiB LLC achieves a geomean speedup of \speedup{\data{1.51}} over 32MiB LLC with no prefetcher.

Halving the LLC from 32MiB to 16MiB reduces \prefetcher{}'s geomean speedup by only \data{7.5\%}, due to a \data{14\%} increase in memory traffic from the higher LLC miss rate.
This reflects the limited spatial locality inherent in graph workloads, which bounds the marginal benefit of additional LLC capacity.

\subsubsection{Impact of LLC Prefetch Delegation}
\label{sec:llc_prefetch_delegation_impact}
LLC prefetch delegation (\S\ref{sec:llc_prefetch_delegation}) trades a minor increase in LLC controller complexity for a significant performance gain.
The optimization improves \prefetcher{} across every algorithm we evaluate, yielding geomean speedups of \speedup{\data{1.12}}, \speedup{\data{1.17}}, \speedup{\data{1.49}}, and \speedup{\data{1.09}} for BC, BFS, PR, and SSSP, respectively, over the non-delegating configuration.
Figure~\ref{fig:data_source_distribution} reveals the underlying mechanism.
By comparing the data-source distributions and access latencies of \emph{default} (delegation enabled) and \emph{no-delegation} configurations, we see that delegation alleviates pressure on the \prefetcher{}Cache, substantially reducing its access latency while keeping LLC access latency close to non-prefetching baseline.

\subsubsection{Impact of Prefetch Drops}
\label{sec:prefetch_drop_impact}
\begin{figure}[t]
    \centering
    \includegraphics[width=\columnwidth]{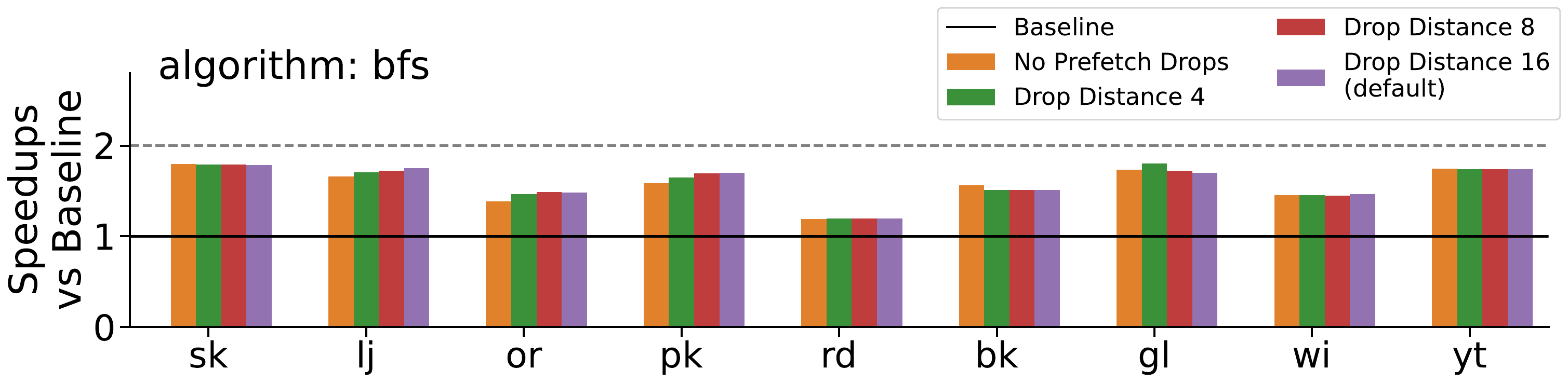}
    \caption{Impact of prefetch drop distance on the performance of \prefetcher{} on BFS.}
    \label{fig:bfs_prefetch_sensitivity}
\end{figure}

Figure~\ref{fig:bfs_prefetch_sensitivity} shows that prefetch dropping improves \prefetcher{} performance on BFS for 4 of 9 evaluated graphs, while the remaining graphs see negligible change.
Relative to no dropping, a drop distance of 16 lifts performance by 5.4\%, 7.9\%, and 7.5\% on the social graphs \emph{livejournal}, \emph{orkut}, and \emph{pokec}, respectively.
These graphs exhibit higher average node degrees and larger active working set sizes, which cause the prefetcher to spend more time traversing each node's neighbors and risk falling behind the core's progress.
A sufficient drop distance filters out requests the core will satisfy imminently, preventing the prefetcher from issuing redundant requests.

\begin{figure}[t]
    \centering
    \includegraphics[width=\columnwidth]{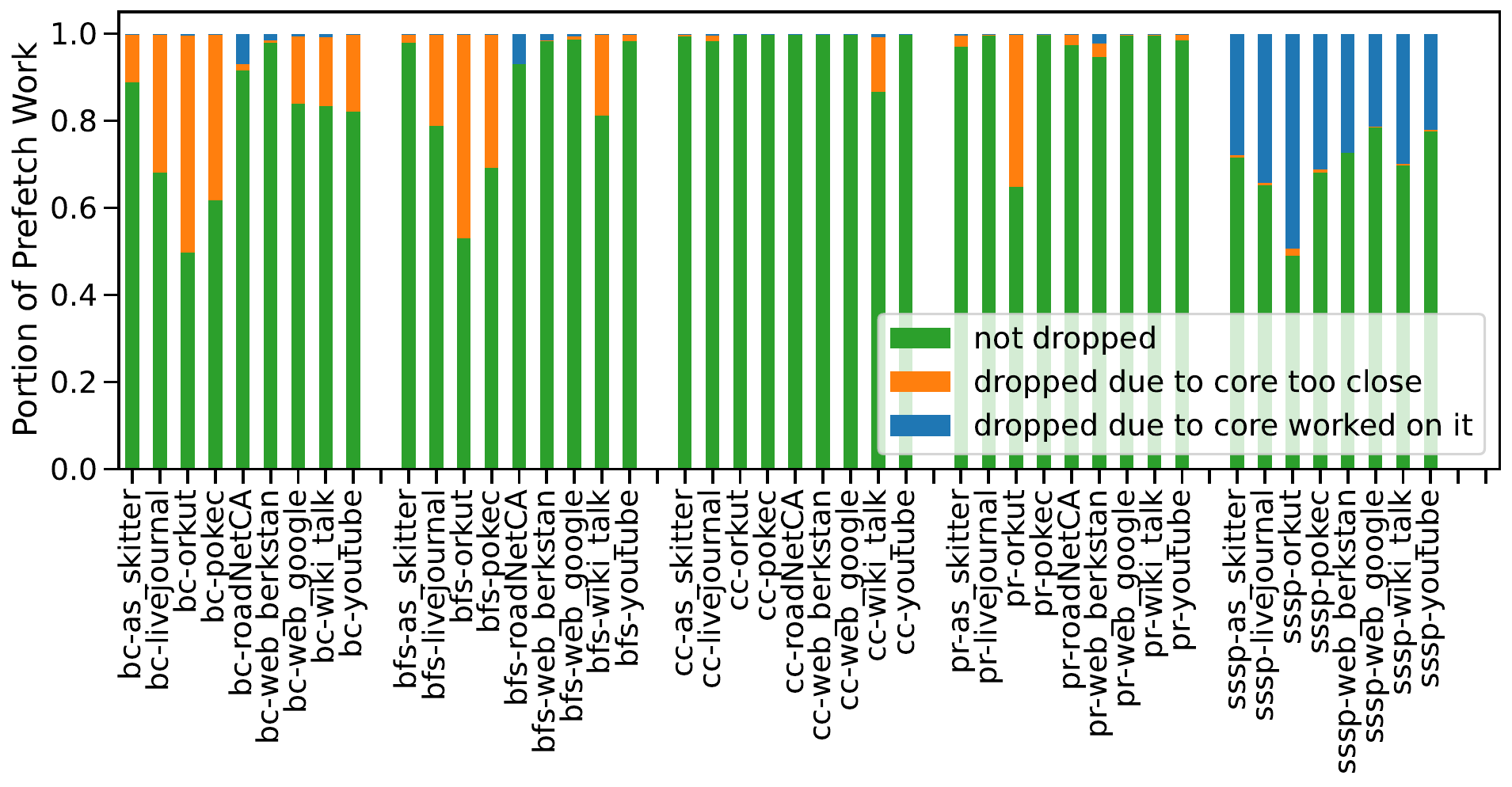}
    \caption{Breakdown of prefetch drop causes for different graph datasets.}
    \label{fig:pickle_prefetch_drop_cause_ratio}
\end{figure}

Figure~\ref{fig:pickle_prefetch_drop_cause_ratio} reflects that a significant amount of prefetch works are dropped, particularly for social graphs where \prefetcher{} traverses high-degree nodes.
The BC, BFS, and PR mainly drops prefetches due to the "too close" cause, suggesting that tuning the drop distance can yield further gains.
On the other hand, the SSSP drops are dominated by the core having already advanced past the target node, indicating that only faster prefetch operations can improve timeliness for this workload.
CC exhibits few drops overall, suggesting that the prefetcher already operates near its full potential on this workload.

\subsubsection{Impact of Prefetching Page Table}

\begin{figure}[t]
    \centering
    \includegraphics[width=\columnwidth]{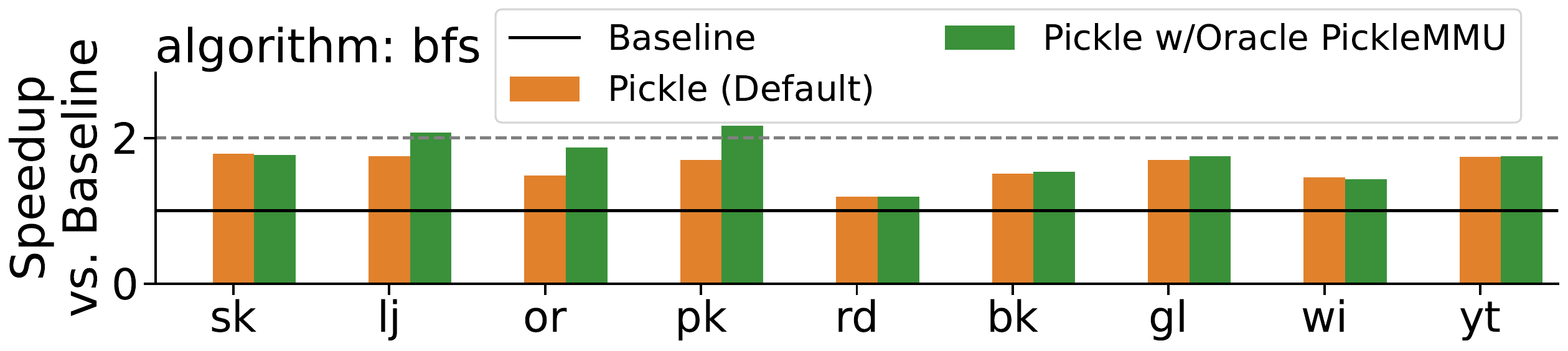}
    \caption{Impact of prefetching page table on the performance of \prefetcher{} on BFS.}
    \label{fig:pickle_functional_mmu_speedup}
\end{figure}

Prefetching using virtual addresses has a side effect of prefetching page table entries (PTEs) along with the data when the prefetcher has a TLB miss.
As the TLB coverage of our evaluated systems is limited compared to the graph size, the core MMUs experience a significant number of TLB misses, which in turn causes the cores to stall.
In this section, we quantify the impact of this side effect on system performance.
In the Oracle PickleMMU system, \prefetcher{} has an oracle that translates all virtual addresses to corresponding physical addresses in a single cycle without changing the state of the system.
In other words, \prefetcher{} only prefetches data and does not trigger page table walks.

Figure~\ref{fig:pickle_functional_mmu_speedup} shows the impact of address translation on the performance of \prefetcher{} on BFS.
Most graphs show similar performance between the two systems, except for the livejournal and pokec graphs, whose speedups degrade from \speedup{2.07} to \speedup{1.74} and \speedup{2.17} to \speedup{1.71}, respectively.
High-degree nodes in these social graphs generate bursts of prefetch requests whose virtual addresses thrash the PickleMMU's TLB, causing the prefetcher to stall on page table walks before it can issue the actual data prefetches.
This is evident in the increase in dropped prefetch requests: from 2.7\% with the oracle PickleMMU to 21.2\% with the PickleMMU for livejournal, and from 0.2\% to 30.9\% for pokec.
This suggests that the performance limitation is mainly due to the bottleneck in address translation.

\subsection{Case Studies on Conditional Prefetching}
\label{sec:case_study}
\subsubsection{Impact of Conditional Branch on Prefetching in SSSP}
\label{sec:case_study_sssp}

\begin{figure}[t]
\begin{lstlisting}[style=cstyle, escapechar=@, backgroundcolor = \color{white},frame=none,lineskip=-5pt, basewidth=0.5em,aboveskip=0pt, belowskip=0pt, numbers=left, numberstyle=\tiny\color{gray}, numbersep=5pt]
while (not finish) {
    distance_threshold = delta * bin_index;  // frontier relaxation threshold
    @\hl{*pickle\_kernel\_3 = distance\_threshold;  // tell pickle the threshold}@
    for (node in current_frontier) {
        @\hl{*pickle\_kernel\_1 = node;  // tell pickle the current frontier }@
        if (node->distance >= distance_threshold) {
            for (neighbor in node->neighbors) {
                // relax neighbor's distance
                ...
            }
        }
    }
    ...
}
\end{lstlisting}
\caption{GAPBS' \cite{beamer2015gap} implementation of delta-stepping loop \cite{meyer2003delta,zhang2020optimizing} for SSSP algorithm. The highlighted lines are the prefetch hints.}
\label{fig:sssp_optimized}
\end{figure}

\begin{figure}[t]
    \centering
    \includegraphics[width=\columnwidth]{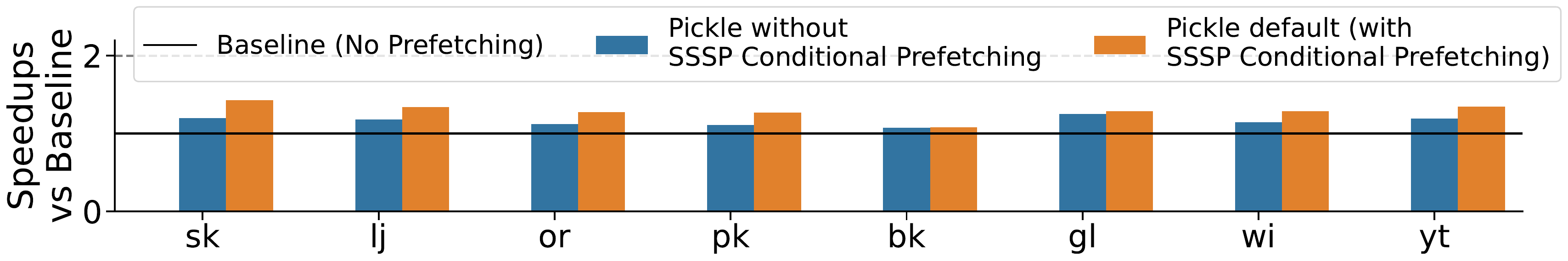}
    \caption{Speedups of \prefetcher{} on SSSP with/without conditional prefetching.}
    \label{fig:pickle_sssp_optimization_speedups}
\end{figure}

\begin{figure}[t]
    \centering
    \includegraphics[width=\columnwidth]{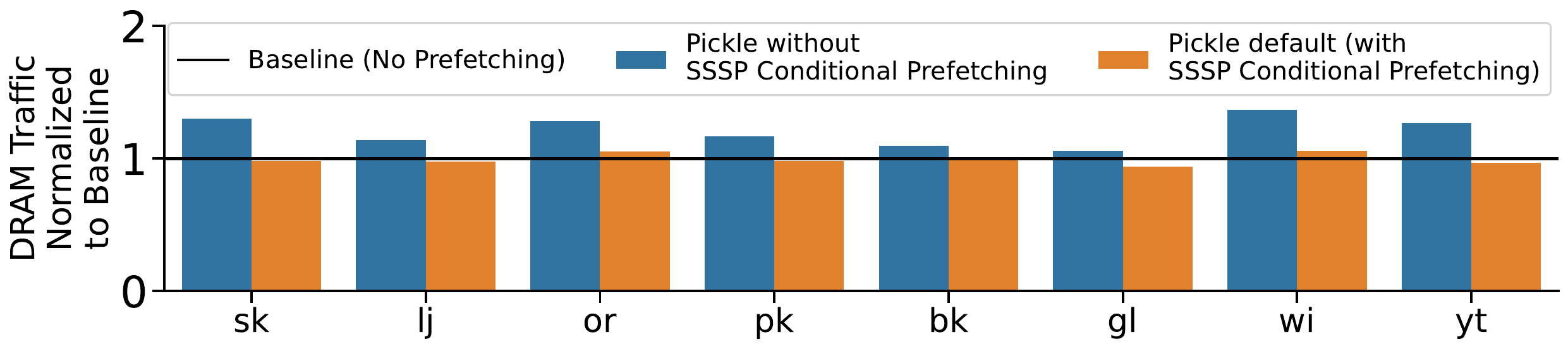}
    \caption{DRAM traffic of \prefetcher{} on SSSP with/without conditional prefetching.}
    \label{fig:pickle_sssp_optimization_dram_bytes}
\end{figure}

In GAPBS' \cite{beamer2015gap} implementation of delta-stepping loop \cite{meyer2003delta,zhang2020optimizing} for SSSP algorithm, delta-stepping only relaxes a neighbor's distance when the current node's tentative distance exceeds a dynamically updated threshold.
Profiling across our graph datasets reveals that this conditional branch is taken 44-84\% of the time, making it a significant source of unnecessary prefetch work when the prefetcher lacks visibility into the branch outcome.

We enable conditional prefetching by communicating the threshold to \prefetcher{}'s context at the start of each delta-stepping iteration: kernel~3 writes the current threshold into the prefetcher context, and during frontier relaxation, kernel~1 checks the stored threshold and drops prefetches for nodes whose distances fall below it.

Figures~\ref{fig:pickle_sssp_optimization_speedups} and~\ref{fig:pickle_sssp_optimization_dram_bytes} quantify the impact.
The optimization raises \prefetcher{}'s geomean speedup on SSSP from \speedup{\data{1.17}} to \speedup{\data{1.29}} and eliminates the DRAM traffic overhead entirely, reducing it from \speedup{\data{1.21}} to \speedup{\data{1.00}}.
This result highlights a key advantage of programmable prefetching: the ability to incorporate high-level algorithmic context into prefetch decisions.

\subsubsection{Impact of Conditional Branch on Prefetching in BC}
\label{sec:case_study_bc}

\begin{figure}[t]
\begin{lstlisting}[style=cstyle, escapechar=@, backgroundcolor = \color{white},frame=none,lineskip=-5pt, basewidth=0.5em,aboveskip=0pt, belowskip=0pt, numbers=left, numberstyle=\tiny\color{gray}, numbersep=5pt]
while (queue not empty) {
    depth++;
    @\hl{*pickle\_kernel\_3 = depth;  // tell pickle the current depth}@
    for (node in queue) {
        @\hl{*pickle\_kernel\_1 = node;  // tell pickle the current frontier }@
        for (neighbor in node->neighbors) {
            if (neighbor->depth == UNASSIGNED) {
                // atomically set neighbor's depth to current depth
                ...
            }
            if (neighbor->depth == depth) {
                // update neighbor's path_counts
                ...
            }
        }
    }
    ...
}
\end{lstlisting}
\caption{GAPBS' \cite{beamer2015gap} implementation of the forward pass of Brandes' algorithm \cite{brandes2001faster,madduri2009faster} for BC. The highlighted lines are the prefetch hints.}
\label{fig:bc_optimized}
\end{figure}

\begin{figure}[t]
    \centering
    \includegraphics[width=\columnwidth]{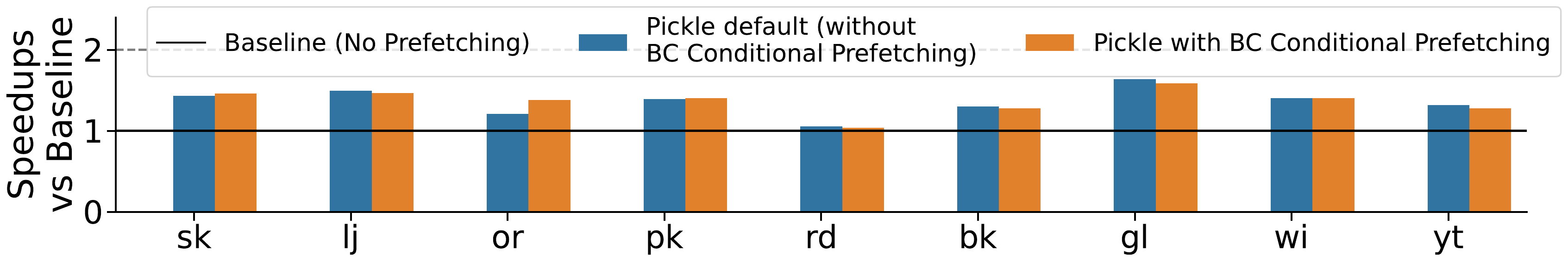}
    \caption{Speedups of \prefetcher{} on BC with/without conditional prefetching.}
    \label{fig:bc_depth_optimization_v1_speedup}
\end{figure}

\begin{figure}[t]
    \centering
    \includegraphics[width=\columnwidth]{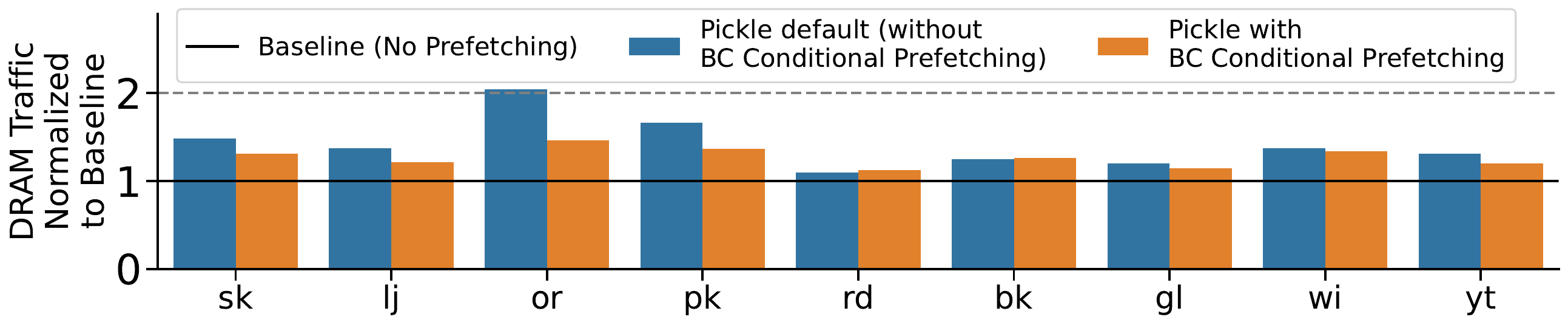}
    \caption{DRAM traffic of \prefetcher{} on BC with/without conditional prefetching.}
    \label{fig:bc_depth_optimization_v1_dram_bytes}
\end{figure}

In GAPBS' \cite{beamer2015gap} implementation of the forward pass of Brandes' algorithm \cite{brandes2001faster,madduri2009faster}, the forward pass of BC updates a neighbor's \texttt{path\_counts} only when the neighbor's depth equals the current BFS level.
This conditional branch is taken about 14\% to 66\% of the time across our graph datasets.

We exploit this by adding a prefetch kernel that communicates the current depth to the prefetcher context.
For neighbors whose depth has not yet been assigned at prefetch time, the prefetcher speculatively assumes assignment at the current depth and prefetches \texttt{path\_counts} accordingly.
In contrast, the non-optimized version unconditionally prefetches both \texttt{depths} and \texttt{path\_counts} for every neighbor, regardless of whether the branch will be taken.

Figures~\ref{fig:bc_depth_optimization_v1_speedup} and~\ref{fig:bc_depth_optimization_v1_dram_bytes} show that, while conditional prefetching does not uniformly improve performance, it substantially reduces DRAM traffic overhead from \speedup{\data{1.40}} to \speedup{\data{1.27}} over the non-prefetching baseline.
The effect is most pronounced on \emph{orkut}, the largest graph in our dataset and the one with the lowest branch-taken rate (14\%): conditional prefetching improves \prefetcher{}'s speedup from \speedup{\data{1.16}} to \speedup{\data{1.36}} and reduces DRAM traffic from \speedup{\data{2.02}} to \speedup{\data{1.47}}.

For the remaining graphs, performance stays roughly flat as two opposing effects counterbalance.
Conditional prefetching requires \prefetcher{} to pull the neighbors' depth values into its own cache to evaluate the branch predicate, forfeiting the LLC delegation optimization (\S\ref{sec:llc_prefetch_delegation_impact}) that the unconditional version utilizes.
However, filtering out unnecessary prefetches reduces \prefetcher{}'s total work volume, freeing bandwidth for higher-value requests.
Therefore, the performance impact of conditional prefetching is highly dependent on the graph structure and the branch-taken rate.

\subsubsection{Impact of Conditional Branch on Prefetching in UA}
\label{sec:case_study_ua}

Each UA mesh-refinement step runs multiple CG iterations, each with a forward and backward heat-transfer phase.
For every element side, a non-conforming side draws contributions from all 100 neighbors, while a conforming side draws from between 29 to 45 neighbors.

We implement conditional prefetching by communicating, per UA step, the current element count to the prefetcher context, and, before each forward heat-transfer call in every CG iteration, the element currently being processed.
The element count is needed because mesh refinement changes the number of elements over time; a prefetch distance exceeding the element count could prefetch out-of-bounds data.
The ua-branch-aware kernel of \prefetcher{} prefetches only the contributing neighbors, whereas the ua-naive kernel unconditionally prefetches all 100 neighbors per side, ignoring conformity.

The ua-branch-aware kernel achieves a speedup of \speedup{\data{2.08}} at \speedup{\data{1.05}} traffic overhead over the non-prefetching baseline, versus \speedup{\data{2.00}} at \speedup{\data{1.11}} for ua-naive; prefetch usefulness is \data{92.6\%} versus \data{83.2\%}.
Because UA's branches are complex, the ua-branch-aware kernel barely fits in the prefetcher's instruction memory, exposing a tradeoff between prefetch accuracy, traffic overhead, and performance.


\subsection{Security Implications}
\label{sec:security_implications}
\prefetcher{} introduces new hardware and software components that interact with the memory subsystem, increasing the surface of side-channel attacks; however, \prefetcher{}'s design choices limit the risk of unauthorized access to thread-private data: the prefetcher is strongly isolated in both hardware and software.
\prefetcher{} operates entirely in userspace and within a single program's virtual address space, so \prefetcher{}'s reach is confined by its MMU.
Moreover, the store requests issued by \prefetcher{}'s RISC-V cores are exclusively routed to the scratchpad, which the program cannot access.
While software can specify how values in the scratchpad drive next-level prefetches, no software API or physical mechanism exists to read scratchpad data back into software.
Consequently, the RISC-V cores' only interaction with the memory subsystem is generating effective addresses for the request manager, leaving no path for unauthorized access to thread-private data.
This leaves only the side-channel attack surface, which can be mitigated by techniques such as randomizing the prefetch order \cite{vreman2019minimizing}.
In practical terms, \prefetcher{} carries a similar security risk to adding more cores to the system.

\subsection{Area/Power Overhead}
\label{sec:area_power_overhead}

%
%

The Snitch RV32E control core \cite{zaruba2020snitch} occupies 10kGE and consumes 1.7mW at 22nm on a 32$\times$32 matrix-multiply benchmark.
Extrapolating to RV64E by doubling the datapath width and register file and adding 50\% place-and-route overhead gives 30kGE and 5.1mW per core.
So a 64-core cluster costs 1.92MGE and 326.4mW in total, or 0.17~mm$^2$ under the ASAP7 NAND2 characteristics \cite{nand2sizeasap7}.

We estimate \prefetcher{}'s SRAM area and power with CACTI~7~\cite{cacti7} at 32nm (Table~\ref{table:SRAM_Overhead}); scaling to 7nm \cite{stillmaker2017scaling} yields 0.35mm$^2$ and 157.4mW.
Combined with the cores, \prefetcher{} adds 0.52mm$^2$ and 483.8mW.
For reference, this corresponds to 0.7\% area and 0.3\% power overhead on an 8-core AMD Zen 5 CCD.

\begin{center}
\footnotesize
\begin{table}
\centering
\caption{SRAM Overhead of \prefetcher{} using 32nm technology}
\begin{tabular}{ | c | c | c | c |}
    \hline
    Component & Description & \makecell{Area \\(mm$^2$)} & \makecell{Power \\(mW)} \\
    \hline
    Prefetch Generator & \makecell{64KiB instruction memory} & 0.1 & 43.6 \\
    \hline
    PickleCache & \makecell{256KiB, 16-way assoc.} & 1.3 & 170.0 \\
    \hline
    TLB & \makecell{L1/L2: 64/1024 entries} & 0.03 & 9.9 \\
    \hline
    Prefetch Context & \makecell{256KiB, Scratchpad} & 0.7 & 153.5 \\
    \hline
    Prefetch Hint Queue & 4.2KiB & 0.2 & 3.0 \\
    \hline
    Request Manager & 32.1KiB & 0.4 & 22.3 \\
    \hline
    Total &  & 2.73 & 402.3 \\
    \hline
\end{tabular}
\label{table:SRAM_Overhead}
\end{table}
\end{center}

\section{Conclusion}
\prefetcher{} lays out the blueprint for a new class of hardware/software co-designed prefetcher that is fully decoupled from the cores and private caches.
Our principle of decreasing the speculativeness of prefetching addresses the wasted bandwidth problem, bringing much-needed predictable prefetcher performance while achieving higher performance than the SOTA IMA data prefetcher with lower memory traffic consumption for irregular workloads.
This shows the potential of this architecture to build latency-tolerant systems.

\begin{acks}
   The writing of this paper is assisted by LLMs.
   We use LLMs to help with grammar and style, but all content and ideas in this paper are our own.
   We take full responsibility for any errors or issues in the paper, including the references.
\end{acks}


\bibliographystyle{ACM-Reference-Format}
\bibliography{references}

\end{document}